\documentclass[preprintnumbers,aps,pra,groupedaddress]{revtex4}
\usepackage{multirow}
\usepackage{hyperref}
\usepackage{mathrsfs}
\usepackage{amsmath}
\usepackage{graphicx}
\usepackage{color}

\hypersetup{
	colorlinks=true,
	linkcolor=blue,
	filecolor=blue,      
	urlcolor=blue,
	citecolor=blue,
}

\def\dsum{\displaystyle\sum}
\begin{document}

\title{Measuring topological invariants for higher-order exceptional points in quantum three-mode systems}

\author{Pei-Rong Han$^{1,2}$}
\author{Wen Ning$^{1}$}
\author{Xin-Jie Huang$^{1}$}
\author{Ri-Hua Zheng$^{1}$}
\author{Shou-Bang Yang$^{1}$}
\author{Fan Wu$^{1}$}
\author{Zhen-Biao Yang$^{1,3}$}
\thanks{E-mail: zbyang@fzu.edu.cn}
\author{Qi-Ping Su$^{4}$}
\author{Chui-Ping Yang$^{4}$}
\thanks{E-mail: yangcp@hznu.edu.cn}
\author{Shi-Biao Zheng$^{1,3}$}
\thanks{E-mail: t96034@fzu.edu.cn}

\address{$^{1}$Fujian Key Laboratory of Quantum Information and Quantum Optics, College of Physics and Information Engineering, Fuzhou University, Fuzhou, China\\
	$^{2}$School of Physics and Mechanical and Electrical Engineering, Longyan University, Longyan, China \\
	$^{3}$Hefei National Laboratory, Hefei, China\\
	$^{4}$School of Physics, Hangzhou Normal University, Hangzhou, China} 

\begin{abstract}
Owing to the presence of exceptional points (EPs), non-Hermitian (NH)
systems can display intriguing topological phenomena without Hermitian
analogs. However, experimental characterizations of exceptional topological
invariants have been restricted to second-order EPs (EP2s) in classical or
semiclassical systems. We here propose an NH \textcolor{black}{multi-mode system} with
higher-order EPs, each of which is underlain by a multifold-degenerate
multipartite entangled eigenstate. We implement the \textcolor{black}{NH model} by
controllably coupling a \textcolor{black}{Josephson-junction-based electronic mode} to two microwave resonators. We
experimentally quantify the topological invariant for an EP3, by mapping out
the complex eigenspectra \textcolor{black}{of the tripartite system along} a loop surrounding this EP3 in the parameter
space. The nonclassicality of the realized topology is manifested by the
observed quantum correlations in the corresponding eigenstates. Our results
extend research of exceptional topology to fully quantum-mechanical models
with multipartite entangled eigenstates. 
\end{abstract}


\narrowtext

\maketitle
\section{Introduction}

As one of the most well-tested physical theories, quantum mechanics has
successfully passed 
\textcolor{black}{numerous experimental tests}. 
In most
quantum-mechanical experiments, the system of interest is well isolated from
its surrounding environment \cite{1,2}, so that its dynamics is governed by the
Schr\"{o}dinger equation with an energy operator often called \textcolor{black}{Hamiltonian}. 
\textcolor{black}{The Hermiticity of the Hamiltonian ensures that the system evolves unitarily.}
However, any real quantum system
\textcolor{black}{is inevitably coupled to the environment}
\cite{3}, which functions by entangling its degrees of
freedom with the \textcolor{black}{system} state \cite{4}. Through this entanglement, the
environment continuously monitors the \textcolor{black}{system} state, which inevitably
produces a measurement backaction \cite{5}. 
\textcolor{black}{Under this disturbance, the \textcolor{black}{system} dynamics could significantly deviate from the unitary evolution even when it does not make any quantum jump.}
Mathematically, the
state trajectory, associated with this conditional evolution, is also
governed by a Schr\"{o}dinger equation but with a non-Hermitian (NH)
Hamiltonian \cite{6}.

Due to the non-Hermiticity, two or more eigenvectors of the NH
Hamiltonian can coalesce into a single one with the same eigenenergy at
exceptional points (EPs) \cite{7,8}. These singularities can bring about many
unique phenomena, exemplified by exotic topological phenomena that are
absent in Hermitian systems \cite{8,9,10,11,12}. The past two decades have witnessed a
number of experimental explorations on the NH singular features, including
spectral parity-time phase transitions \cite{13,14,15,16,17,18,19,20,21}, dynamical chiral behaviors
\cite{22,23,24,25,26,27,28}, exceptional entanglement transitions \cite{29}, and NH topology \cite{30,31,32,33,34,35,36,37,38,39,40,41,42,43,44}.
\textcolor{black}{The topological invariant
of an NH system can be quantified in terms of the eigenvectors or the complex eigenenergies \cite{8,9,10,11,12}. This is in distinct contrast with the Hermitian case, where the topology cannot be defined by the eigenenergies which are always real.} 
So far, \textcolor{black}{NH topology has been observed} in several experiments,
all of which were restricted to second-order EPs (EP2s) realized in
classical systems \cite{35,36,37,38,39} or with a classically-driven qubit \cite{40,41,42}.
Compared to EP2s, higher-order EPs can exhibit much richer topological
properties \cite{11,12,43,44,45} and the associated spectral topological
invariants are defined in a fundamentally different manner \cite{12}.
Despite fundamental interest, such invariants have not been unambiguously
characterized in experiment.

\textcolor{black}{We here investigate both theoretically and experimentally the quantum-mechanical exceptional topology associated with higher-order EPs in an NH composite system consisting of multiple interacting modes. These EPs result from continuous and nonunitary evolutions in the single-excitation subspace without quantum jumps \cite{50}.}
The topological property associated to each
higher-order EP can be quantified by the homotopy invariant, recently
proposed by Delplace et al. \cite{12}. We experimentally engineer the NH
\textcolor{black}{three-mode system} in a superconducting circuit, where \textcolor{black}{a nonlinear Xmon mode} is
controllably coupled to two photonic modes stored in two separated
microwave resonators. 
\textcolor{black}{\textcolor{black}{The non-Hermiticity of the system} is manifested by the non-negligible dissipation of one photonic mode.}
We find that the resulting NH Hamiltonian possesses four
EP3s in the real parameter space. We quantify the homotopy invariant
associated to the EP3 in the first quadrant of the parameter space, by
mapping out the eigenenergies of the NH Hamiltonian along a loop
surrounding this EP3. 
\textcolor{black}{The system eigenenergies are extracted from the output states of the three-mode system in the single-excitation subspace, measured for different evolution times. As far as we know, this is the first experimental characterization of the topological invariant associated with a higher-order EP.}
\textcolor{black}{We further note that the winding number defined by Eq. (\ref{Eq3}) is zero for the degenerate points in both the two-dimensional (2D) and 3D Hermitian systems, referred to as diabolical points (DPs), as well as for EP2s in a 2D NH system, as detailed in the Supplemental Material. Therefore, such a winding number serves as a homotopy invariant that uniquely characterizes the topology of EP3s.}

\section{Results}
\subsection{NH multi-mode system}
\color{black}
The system under consideration corresponds to a multi-mode system with competing coherent nearest-neighboring interactions and incoherent dissipation, as shown in Fig. 1a. Under the competition, the system evolution is a weighted mixture of infinitely many trajectories, among which of special interest is the one without quantum jump. This trajectory is governed by the NH Hamiltonian (setting $\hbar =1$) 
\begin{equation}
H_{\mathrm{NH}}=-\frac{1}{2}i\dsum\limits_{j=1}^{N}\kappa _{j}a^{\dagger}_j a_j
+\dsum\limits_{j=1}^{N-1}\lambda _{j}(a _{j}^{\dagger}a _{j+1}+\mathrm{H.c.}),
\label{NH H}
\end{equation}%
where $a _{j}^{\dagger}$ and $a_j$ denote the creation and annihilation operators of the $j$th mode with a decaying rate $\kappa _{j}$, $\lambda _{j}$ represents the coupling coefficient between the $j$th and ($j+1$)th modes\textcolor{black}{, and H.c. is the Hermitian conjugate}. The excitation number of the total system is conserved along the no-jump trajectory, as the \textcolor{black}{excitation number} operator, $N_e = \sum_{j=1}^{N} a_{j}^{\dagger}a_{j}$, commutes with $H_{\mathrm{NH}}$. Hereafter, we will consider the system behaviors restricted in the single-excitation subspace 
$\left\{ |1_1 0_2...0_N\rangle ,..., |0_1...0_{N-1}1_N \rangle \right\}$, 
where $|0_j\rangle$ and $|1_j\rangle$ respectively denote the ground and first excited states for the $j$th mode with $j$ \textcolor{black}{ranging from $1$ to $N$.}
\color{black}

\begin{figure}[htbp]
	\centering
	\includegraphics[width=3.3in]{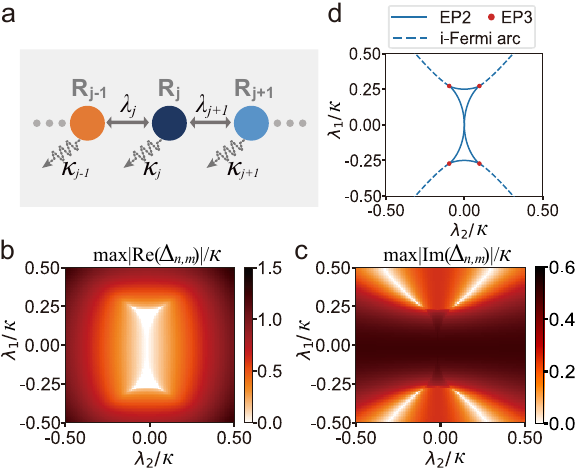}
	\caption{\textcolor{black}{Sketch of the NH multimode system and the spectral structure. (a) Theoretical model. The system involves $N$ modes ($R_{j}$), arranged in a linear array.} The NH Hamiltonian dynamics features
		the competition between the nearest-neighbor swapping couplings ($\protect%
		\lambda _{j}$) and the energy dissipations ($\protect\kappa _{j}$). (b), (c)
		Maxima for the real (b) and imaginary (c) gaps versus $\protect\lambda _{1}$
		and $\protect\lambda _{2}$ in the \textcolor{black}{three-mode system}. For simplicity, we set $%
		\protect\kappa _{2}=\protect\kappa _{3}=0$, and assume $\protect\kappa _{1}$
		has a fixed nonzero value $\protect\kappa $. At the three-fold degeneracy
		points $\left\{ \pm \protect\lambda _{1}^{c},\pm \protect\lambda %
		_{2}^{c}\right\}$, the maximum gap among the three complex eigenenergies
		vanishes. Here the parameters $\protect\lambda _{1}$ and $\protect\lambda %
		_{2}$ are scaled in unit of $\protect\kappa $. $\Delta_{n,m}$ denotes the difference
		between two eigenenergies, i.e., $E_n-E_m$ ($n\neq m$, $n,m=1,2,3$).
		(d) The solid lines denote EP2s where a pair of eigenstates coalesce.
		Along the dashed lines, all the three eigenenergies have the same
		imaginary part. Their intersection points (red dots) are EP3s connecting Fermi
		arcs and i-Fermi arcs.}
	\label{Fig1}
	\vspace*{-0.15in}
\end{figure}

For $\kappa _{j}=0$ ( $j=1,...,N$), the Hamiltonian (\ref{NH H}) has $N$%
\ real eigenenergies, which display an $N$-fold degeneracy for $\lambda
_{j}=0$. When one or more \textcolor{black}{modes} are subjected to dissipations, 
the eigenenergies become complex. The resulting spectral structure
can display exotic topological features without Hermitian analogues. To
clearly illustrate the underlying NH quantum topological physics, we
consider a \textcolor{black}{three-mode} system. For simplicity, we assume $\kappa _{2}=\kappa
_{3}=0$, and drop off the subscript of $\kappa _{1}$. Fig. \ref{Fig1}b (c) shows
the maximal real (imaginary) part of the gaps among the three complex
eigenenergies versus $\lambda _{1}$\ and $\lambda _{2}$. The results clearly
show that the non-Hermiticity splits the real degeneracy into four EP$3$s at 
$\left\{ \pm \lambda _{1}^{c},\pm \lambda _{2}^{c}\right\} $ with $\lambda
_{1}^{c}=\sqrt{2}\kappa /3\sqrt{3}$ and $\lambda _{2}^{c}=\kappa /6\sqrt{3}$%
. At each EP3, the three eigenenergies have a vanishing real part and the
same nonzero imaginary part.

The spectrum of the NH Hamiltonian exhibits a rich structure,
where each of the four EP3s is connected to two curves of EP2s (solid lines
in Fig. \ref{Fig1}d), at which a pair of eigenstates and eigenenergies coalesce. The
details of these eigenenergies are shown in \textcolor{black}{Sec. S5 of the} Supplemental Material. Enclosed
in the EP2 contour is an isofrequency region, where all the three
eigenenergies have the same real part. This is a 2D generalization of the
linear-like two-fold degenerate Fermi arc, which connects 2 EP2s 
existing in a 2D system \cite{30}. 
\textcolor{black}{In the isofrequency region, the real parts of the three complex eigenenergies} \textcolor{black}{vanish}, \textcolor{black}{so that} \textcolor{black}{the state evolution of the system} \textcolor{black}{is determined by the imaginary parts of these eigenenergies, each of which corresponds to a gaining or losing rate depending upon its sign. After a long-time evolution, only the eigenvector with the largest imaginary part can survive for the no-jump case. Since such an eigenvector is essentially a tripartite entangled state, the dynamics in the isofrequency region provides a way for robust generation and conditional stabilization of tripartite entanglement \cite{51}. 
Such an entanglement generation and stabilization process is illustrated by the simulations presented in Sec. S2 of the Supplemental Material.
}

In addition to the real Fermi arc, each of
the four EP3s emanates an i-Fermi arc (dashed lines in Fig. \ref{Fig1}d) \cite{11}, along
which the imaginary parts are three-fold degenerate. With a global shift of
the spectrum by a suitable imaginary value, the four i-Fermi arcs correspond
to the regions where the eigenenergies are all real. In view of spectral
phase transitions, each of the four EP3s corresponds to a tripoint,
where three distinct phases (featuring imaginary, real, and complex energy
gaps) meet together.

\subsection{Experimental implementation}
\begin{figure}[htbp]
	\centering
	\includegraphics[width=3.5in]{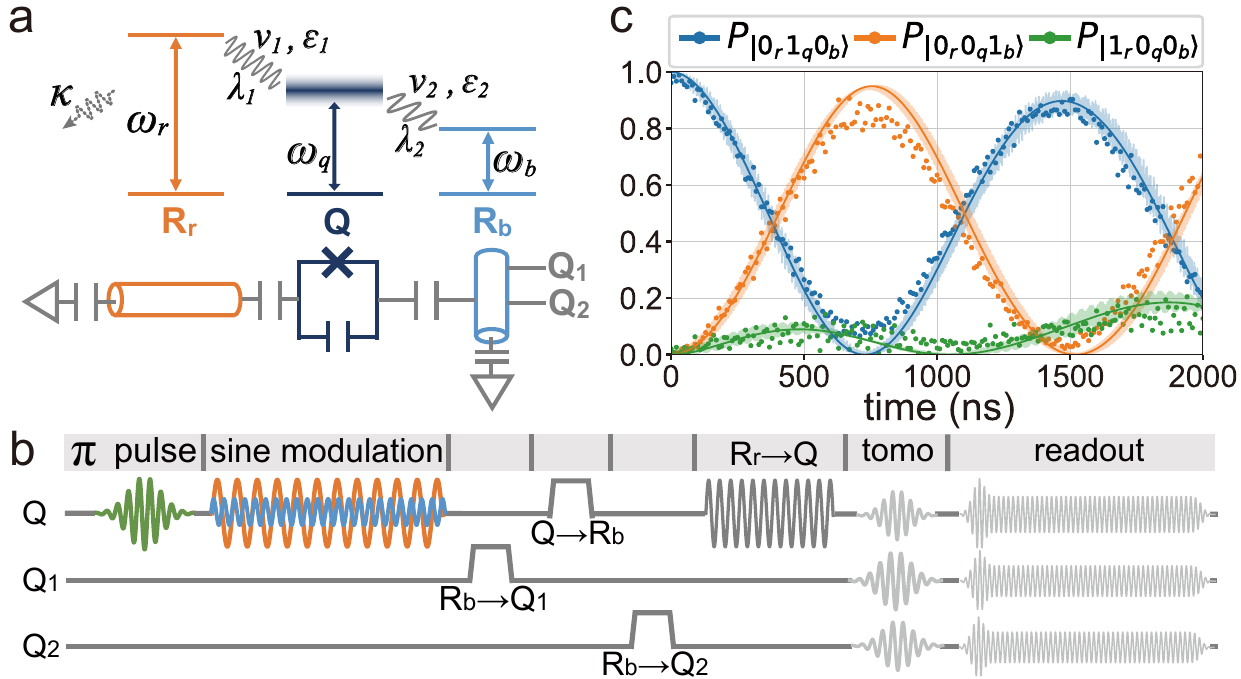}
	\caption{\textcolor{black}{Experimental implementation. (a) Synthesis of the NH three-mode system.} The NH Hamiltonian is realized in a
		circuit, where a Josephson-junction-based qubit ($Q$), together with a bus
		resonator ($R_{b}$) and a readout resonator ($R_{r}$), comprises an
		effective \textcolor{black}{three-mode system} in the single-excitation subspace. The decaying
		rates of $Q$ and $R_{b}$ are respectively 0.06 MHz and 0.08 MHz, both of
		which can be neglected compared to that of $R_{r}$, $\protect\kappa =5$ MHz. 
		$Q$ is coupled to $R_{r}$ ($R_{b}$) at the first upper (lower) sideband with
		respect to the first (second) parametric modulation. (b) Pulse sequence.
		The qubit $Q$ is first prepared in the excited state at its idle frequency,
		followed by the application of two sine modulations. The modulation
		frequencies ($\protect\nu _{1}$, $\protect\nu _{2}$) and amplitudes are
		tunable for controlling $\protect\lambda _{1}$ and $\protect\lambda _{2}$.
		After the modulating pulse, the evolved \textcolor{black}{$R_{r}$-$Q$-$R_{b}$} output state is
		mapped to the \textcolor{black}{$Q$-$Q_{2}$-$Q_1$} system for readout, where $Q_{1}$ and $%
		Q_{2}$ are two ancilla qubits, each of which can be controllably coupled to
		the bus resonator. (c) Observed evolutions of the populations. All data are
		measured for the initial state \textcolor{black}{$\left\vert 0_{r}1_{q}0_{b}\right\rangle $} at
		the point $\lambda _{1}= 2\pi\times 0.21$ MHz and $\lambda
		_{2}=2\pi\times0.31$ MHz. The results are obtained by discarding the
		measurement outcome \textcolor{black}{$\left\vert 0_{r}0_{q}0_{b}\right\rangle $} and
		renormalizing the remaining populations of \textcolor{black}{$\left\vert
			0_{r}0_{q}1_{b}\right\rangle ,$ $\left\vert 0_{r}1_{q}0_{b}\right\rangle $,
			and $\left\vert 1_{r}0_{q}0_{b}\right\rangle $} in the single-excitation
		subspace. The solid curves are theoretical predictions using the NH
		Hamiltonian (\ref{NH H}), while the fast oscillating curves are
		numerical simulation results using the original Hamiltonian (given in \textcolor{black}{Sec. S3 of the}
		Supplemental material) with frequency modulations included.}
	\label{Fig2}
\end{figure}

We perform the experiment using a circuit quantum electrodynamics device
with five \textcolor{black}{nonlinear Xmon modes (qubits)}, each controllably coupled to the common bus resonator
($R_{b}$) of a fixed frequency $\omega _{b}/2\pi=5.58 $ GHz. The NH quantum multipartite system is realized with one of these qubits
(labeled as $Q$), the bus resonator $R_{b}$, and \textcolor{black}{the readout resonator of $Q$, ($R_{r}$)}, which has
a frequency $\omega _{r}/2\pi=6.66$ GHz. The on-resonance
coupling strength between $Q$ and $R_{b}$ ($R_{r}$) is $g_{b}=2\pi\times20$ MHz (%
$g_{r} =2\pi\times41$ MHz). The decaying rate of $Q$ ($R_{b}$) is $0.06$ ($0.08$) MHz, which is two orders smaller than that of $R_{r}$ ($%
\kappa =5$ MHz). Thus, the dissipations of $Q$ and $R_{b}$ are
negligible.
\textcolor{black}{To realize the NH tripartite Hamiltonian in a
	controllable manner, two parametric modulations are simultaneously applied
	to $Q$ to modulate its transition frequency as $\omega _{q}=\omega
	_{0}+\sum_{j=1,2}\varepsilon _{j}\cos (\nu _{j}t)$, where $\omega
	_{0}$ is the mean frequency of $Q$, and $\varepsilon _{j}$ and $\nu _{j}$
	respectively denote the amplitude and frequency of the $j$th modulation, as
	depicted in Fig.~\ref{Fig2}a. Under the condition $\omega _{0}+\nu _{1}=\omega _{r}$,
	the qubit $Q$ interacts with the readout resonator $R_{r}$
	at the first upper sideband for the first modulation, with the strength $%
	\lambda _{1}=g_{r}J_{1}(\varepsilon _{1}/\nu _{1})$. Here, $J_{1}(\mu )$ is
	the first-order Bessel function of the first kind.\ The second modulation is
	used to induce the sideband interaction between $Q$ and $R_{b}$, with the
	effective photonic swapping rate $\lambda _{2}$\ controlled by $\varepsilon
	_{2}$. The sideband interactions and the system parameters are detailed in Sec. S3 of the Supplemental Material.}
\textcolor{black}{When the $R_r$-$Q$-$R_b$ system initially has one excitation, its dynamics,} 
\textcolor{black}{associated with no-jump trajectory, is described by the NH Hamiltonian (\ref{NH H}) with $N=3$, where $R_r$, $Q$, and $R_b$ correspond to three modes sharing a single excitation.}

The experiment starts with the preparation of an initial single-excitation
state, following which the parametric modulations are applied to $Q$, with
the experimental pulse \textcolor{black}{sequence} depicted in \textcolor{black}{Fig.~\ref{Fig2}b}. 
After a preset interaction time $t$, the parametric modulations are
switched off to have $Q$ decoupled from
both $R_{b}$ and $R_{r}$.
\textcolor{black}{The output state of the system, associated with the no-jump trajectory, is given by}
\begin{equation}
\left\vert \psi (t)\right\rangle =c_{1}(t)\left\vert
0_{r}0_{q}1_{b}\right\rangle +c_{2}(t)\left\vert
0_{r}1_{q}0_{b}\right\rangle +c_{3}(t)\left\vert
1_{r}0_{q}0_{b}\right\rangle .
\end{equation}%
Here the subscripts $r$, $q$, and $b$
label $R_{r}$, $Q$, and $R_{b}$, respectively.
\textcolor{black}{As the excitation is conserved by the NH Hamiltonian while the quantum jump breaks down this conservation, the no-jump state trajectory can be postselected by discarding the outcome with null excitation.}
The joint tripartite output state is read out with the assistance of two
ancilla qubits, denoted as $Q_{1}$ and $Q_{2}$. The \textcolor{black}{$R_{r}$-$Q$-$R_{b}$}
output state is mapped to the \textcolor{black}{$Q$-$Q_{2}$-$Q_{1}$} system, which is realized
by swapping gates (see \textcolor{black}{Sec. S4 in the} Supplemental Material). Bloch vectors of the three
qubits along different axes are then measured. By correlating the outcomes
of these measurements, the three-qubit density matrix is reconstructed.
Removing the ground state element and renormalizing the remaining ones, we
obtain the final state of the system evolving under the NH Hamiltonian.
With a correction for the infidelity of the state mapping, the
resulting \textcolor{black}{$Q$-$Q_{2}$-$Q_{1}$} output state corresponds to the \textcolor{black}{$R_{r}$-$Q$-$R_{b}$} output state right before the state mapping. 
\textcolor{black}{Fig.~\ref{Fig2}c shows the evolution of the populations \textcolor{black}{$\left\vert 0_{r}0_{q}1_{b}\right\rangle $, $\left\vert 0_{r}1_{q}0_{b}\right\rangle $, and $\left\vert
	1_{r}0_{q}0_{b}\right\rangle $ for the initial state $\left\vert
	0_{r}1_{q}0_{b}\right\rangle $}, measured at the point $\lambda _{1}
	=2\pi\times0.21$ MHz and $\lambda _{2}=2\pi\times0.31$ MHz. These populations are obtained
	by discarding the measurement outcome \textcolor{black}{$\left\vert
	0_{r}0_{q}0_{b}\right\rangle $} and then renormalizing the probabilities for
	occurrences of the three single-excitation outcomes.}

\subsection{Measurement of the winding number}
\begin{figure*}[htbp]
	\centering
	\includegraphics[width=6in]{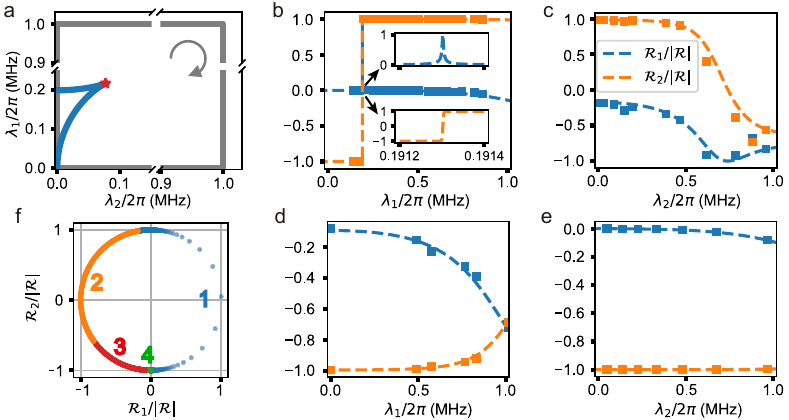}
	\caption{Characterization of the NH topology. (a)
		Loop traversed for extracting the winding number. The loop (gray solid line)
		is square-shaped with four vertices ($0,0$), ($\protect\lambda _{m},0$), ($0,%
		\protect\lambda _{m}$), and ($\protect\lambda _{m},\protect\lambda _{m}$),
		where $\lambda _{m}=2\pi\times1$ MHz. This loop surrounds an EP3
		(red star), which connects two Fermi arcs consisting of EP2s (blue
		solid line). The arrow denotes the direction of the integral along the loop.
		(b)-(e) Normalized resultant vectors measured against $\protect\lambda %
		_{1} $ with $\protect\lambda _{2}=0$ (b) and $\protect\lambda_{m}$ (d), and
		against $\protect\lambda _{2}$ with $\protect\lambda _{1}=\lambda_{m}$ (c) and $0$ (e). The squares represent the components $\mathcal{R}%
		_{1}/|\mathcal{R}|$ (blue) and $\mathcal{R}_{2}/|\mathcal{R}|$ (orange),
		respectively. The dotted lines denote the fitted $\mathcal{R}_{1}/|\mathcal{R%
		}|$ and $\mathcal{R}_{2}/|\mathcal{R}|$ as functions of $\protect\lambda %
		_{1} $ or $\protect\lambda _{2}$. The winding number, calculated with these
		fitted functions, is $1$. (f) Trajectory of the rescaled unit resultant vector $\mathcal{R}^{u}=(\mathcal{R}_1+i\mathcal{R}_2)/|\mathcal{R}|$. The symbols ``1", ``2", ``3", and ``4", label the sections associated with the four edges of the parameter-space loop, shown in (b), (c), (d), and (e), respectively.}
	\label{Fig3}
\end{figure*}

The topological invariant associated with each EP3 can be quantified by the
winding number, calculated in terms of the resultant vector $\mathcal{R}$
\cite{12} 
\begin{equation}
\mathcal{W}=\frac{1}{2\pi}\dsum\limits_{j=1,2}\oint\limits_{C_{\lambda }}\frac{1}{
\left\Vert \mathcal{R}\right\Vert ^{2}}\left( \mathcal{R}_{1}\frac{\partial 
\mathcal{R}_{2}}{\partial \lambda _{j}}-\mathcal{R}_{2}\frac{\partial 
\mathcal{R}_{1}}{\partial \lambda _{j}}\right) d\lambda _{j},
\label{Eq3}
\end{equation}%
where the integral loop $C_{\lambda }$ encloses the EP3 in the 2D parameter
space. $\mathcal{R}_{1}$ and $\mathcal{R}_{2}$ depend upon the
eigenenergies (for details, see \textcolor{black}{Sec. S7 in the} Supplemental Material), which can in
principle be extracted from the output states measured for different
interaction times. To simplify the measurement, we choose a square-shaped
loop on the $\lambda _{1}$-$\lambda _{2}$ plane. The loop has four
vertices ($0,0$), ($\lambda _{m},0$), ($0,\lambda _{m}$), and ($\lambda
_{m},\lambda _{m}$), as shown in Fig. \ref{Fig3}a. 
\color{black}
With this choice, the eigenspectrum can be extracted in a relatively easy manner 
(see \textcolor{black}{Sec. S5 in the} Supplemental Material). In our experiment, $\lambda_{m}$ is set to
be $2\pi\times1$ MHz.
\color{black}

The eigenenergies, measured along the four edges of the square-shaped loop,
are detailed in \textcolor{black}{Sec. S8 of the} Supplemental Material. For different values of $%
\lambda _{1}$ or $\lambda _{2},$ $\mathcal{R}_{1}$, 
and $\mathcal{R}_{2}$ are calculated from the corresponding
measured eigenenergies, as shown in Fig. \ref{Fig3}b-e. The dashed lines denote the
functions fitted with the measured data. 
We note that the rescaled unit resultant vector, defined as $\mathcal{R}^{u}=(\mathcal{R}_1+i\mathcal{R}_2)/|\mathcal{R}|$, is changed along a circle when the control parameter $(\lambda_1,\lambda_2)$ is varied along a loop enclosing the EP3, as illustrated in Fig. \ref{Fig3}f. On the edge with $\lambda_2=0$, there is an EP2, within the vicinity of which the resultant vector undergoes a $\pi$ rotation, as detailed in the insets of Fig. \ref{Fig3}b. This rotation occurs in a very narrow region of the control parameter $(\lambda_1,\lambda_2)$, but contributes half of the circular trajectory of $\mathcal{R}^{u}$.
The winding number based on
thus-obtained $\mathcal{R}_{1}$ and $\mathcal{R}_{2}$ is $\mathcal{W}=1$, which
confirms that the loop encircles an EP3. In addition to the EP3, the loop
surrounds infinitely multiple EP2s, which form two exceptional
arcs that cross at the EP3, as illustrated in Fig. \ref{Fig3}a. These EP2s have no
contribution to the characterized topological invariant, which
corresponds to the winding number of the relative angle between $\mathcal{R}%
_{1}$\ and $\mathcal{R}_{2}$ \cite{12}. This implies that the spectral
topological features of higher EPs are fundamentally distinct from those of
EP2s. As detailed in Ref. \cite{12}, the winding number defined in Eq. (\ref{Eq3}) is the
simplest example of the homotopy invariants, which can quantify
topological properties of multifold symmetry-protected EPs but have not
been experimentally characterized so far.

In a recent acoustic experiment \cite{43}, the measured local phase rigidities of the eigenvectors near an EP3
have a critical exponent coinciding with the winding number quantifying the
calculated global Berry phase. However, the Berry phase itself,
which characterizes the topological charge associated to the eigenvectors,
has not been experimentally extracted, neither has the eigenspectral
topological invariant. In a recent experiment \cite{49}, a third-order exceptional line was observed with a
nitrogen-vacancy system, where the NH Hamiltonian for a single electron with
three levels was constructed by the dilation method. However, the
topological properties of the realized EP3 \textcolor{black}{have} not been experimentally
characterized.
As far as we know, our work represents the first measurement of homotopy invariants, of which the winding number for EP3s is the simplest example \cite{12}.


\textcolor{black}{To reveal the quantum correlations among the three modes, we perform quantum state tomography on the output state of the system.}
For different values of the control parameters, the
evolutions of the measured pairwise concurrences are detailed in
\textcolor{black}{Sec. S5 of the} Supplemental Material. For the edges $\lambda _{1}=0$\ and $\lambda _{2}=0$,
the system reduces to a \textcolor{black}{two-mode system}, so that the eigenstates can be
extracted from the state evolution in a relatively easy way. At point ($0,\lambda _{m}$), the measured concurrences of the two eigenstates (%
$\left\vert \Phi _{\pm }\right\rangle $) for the $R_{b}$-$%
Q $ subsystem, which is decoupled from $R_{r}$, are 0.997 and 0.997,
respectively. At point ($\lambda _{m},0$), the concurrences for two measured 
$Q$-$R_{r}$ eigenstates ($\left\vert \Phi _{\pm }^{\prime }\right\rangle $)
are respectively 0.971 and 0.971. When $\lambda _{1}\neq 0$ and $\lambda _{2}\neq 0$%
, each of the three \textcolor{black}{eigenstates corresponds to} a tripartite entangled
eigenstate, for which each \textcolor{black}{mode} is quantum-mechanically correlated to
either of the other two \textcolor{black}{modes}. Experimental extraction of these
eigenstates is a challenging task because each eigenstate involves four
parameters to be determined. However, the observation of tripartite
entangled states, evolved from the initial product state $\left\vert
0_{b}1_{q}0_{r}\right\rangle $, indicate$s$ that there exists tripartite
entanglement in the underlying eigenstates. These observed nonclassical
correlations represent another unique feature that fundamentally
distinguishes the presently observed topology from those previously
demonstrated in classical or semiclassical systems \cite{30,31,32,33,34,35,36,37,38,39,40,41,42,43,44,45}.

\section{Discussion}
In conclusion, we have investigated the exceptional topology in interacting
many-body quantum systems, with a competition between coherent
couplings and incoherent dissipation. The quantum topology is manifested by
the presence of multiple higher-order EPs, each of which carries a quantized
topological charge and is associated with a multi-fold degenerate eigenstate
displaying highly nonclassical correlations. The NH model is experimentally
realized with a superconducting qubit, which is correlated to a lossless bus
resonator and a decaying readout resonator by swapping a single photon. The
topological charge at each EP3 is quantified by the winding number,
extracted from the eigenspectra measured with the assistance of two ancilla
qubits. Each of the corresponding eigenstates exhibits quantum entanglement,
confirming the nonclassical origin of the topology.
\textcolor{black}{Besides fundamental interest, the demonstrated NH dynamics associated with higher order EPs may have applications in quantum technologies, such as sensitivity enhancement in quantum metrology \cite{52} as well as fast and robust generation of quantum entanglement \cite{53}.}

\section{Data Availability}
The data that support the findings of this study are available from the corresponding author upon request.

\section{Acknowledgements}
S. B. Zheng was supported by the National Natural Science Foundation of China (Grant Nos. 12274080, 12474356) and the Innovation Program for Quantum Science and Technology (Grant No. 2021ZD0300200); Z. B. Yang was supported by the National Natural Science Foundation of China (Grant No. 12475015) and the Innovation Program for Quantum Science and Technology (Grant Nos. 2021ZD0300200); C. P. Yang was supported by the National Natural Science Foundation of China (Grant No. U21A20436) and the Innovation Program for Quantum Science and Technology (Grant Nos. 2021ZD0301705).

\section{Author Contributions}
S-.B.Z. proposed the theoretical model and conceived the experiment. P-.R.H., W.N., X-.J.H. supervised by Z-.B.Y. and S-.B.Z., carried out the experiment. P-.R.H., R-.H.Z, S-.B.Y, F.W., Q-.P.S, C-.P.Y., and S-.B.Z. analyzed the data. S-.B.Z., Z-.B.Y., and C-.P.Y. cowrote the paper. All authors contributed to interpretation of observed phenomena, and helped to write the paper.

\section{Competing interests}
The authors declare no competing interests.

\newpage

\newpage

\newpage

\end{document}


\title{Supplementary Materials for \protect\\``Measuring topological invariants for higher-order exceptional points in quantum three-mode systems"}

\author{Pei-Rong Han$^{1,2}$}
\author{Wen Ning$^{1}$}
\author{Xin-Jie Huang$^{1}$}
\author{Ri-Hua Zheng$^{1}$}
\author{Shou-Bang Yang$^{1}$}
\author{Fan Wu$^{1}$}
\author{Zhen-Biao Yang$^{1,3}$}
\thanks{E-mail: zbyang@fzu.edu.cn}
\author{Qi-Ping Su$^{4}$}
\author{Chui-Ping Yang$^{4}$}
\thanks{E-mail: yangcp@hznu.edu.cn}
\author{Shi-Biao Zheng$^{1,3}$}
\thanks{E-mail: t96034@fzu.edu.cn}

\address{$^{1}$Fujian Key Laboratory of Quantum Information and Quantum Optics, College of Physics and Information Engineering, Fuzhou University, Fuzhou, China\\
	$^{2}$School of Physics and Mechanical and Electrical Engineering, Longyan University, Longyan, China \\
	$^{3}$Hefei National Laboratory, Hefei, China\\
	$^{4}$School of Physics, Hangzhou Normal University, Hangzhou, China} 

\vskip0.5cm

\narrowtext
\maketitle

\tableofcontents

\section{Eigenenergies and eigenstates of the NH three-mode system}

The \textcolor{black}{non-Hermitian (NH)} model under consideration involves three \textcolor{black}{modes} with nearest-neighbor
couplings. The first \textcolor{black}{mode} has a non-negligible dissipation rate $\kappa $,
while dissipation rates of the \textcolor{black}{other} two \textcolor{black}{modes} are negligible. The system
dynamics associated with the no-jump trajectory is governed by the NH
Hamiltonian (hereafter  setting $\hbar =1$)
\color{black}
\begin{equation}
H=-\frac{1}{2}i\kappa a_{1}^{\dagger}a_{1} +\left(\lambda _{1}a_{1}^{\dagger}a_{2}+\lambda
_{2}a_{2}^{\dagger}a_{3}+\text{H.c.}\right), 
\label{H_NH}
\end{equation}
where $a_{j}^{\dagger}$ and $a_{j}$ ($j=1,...,3$) denote the creation and annihilation operators of the $j$th mode, $\lambda_j$ is the coupling coefficient between the $j$th and $(j+1)$th modes, and H.c. indicates the Hermitian conjugate.
\color{black}

\begin{figure*}[htbp]
	\centering
	\includegraphics[width=6in]{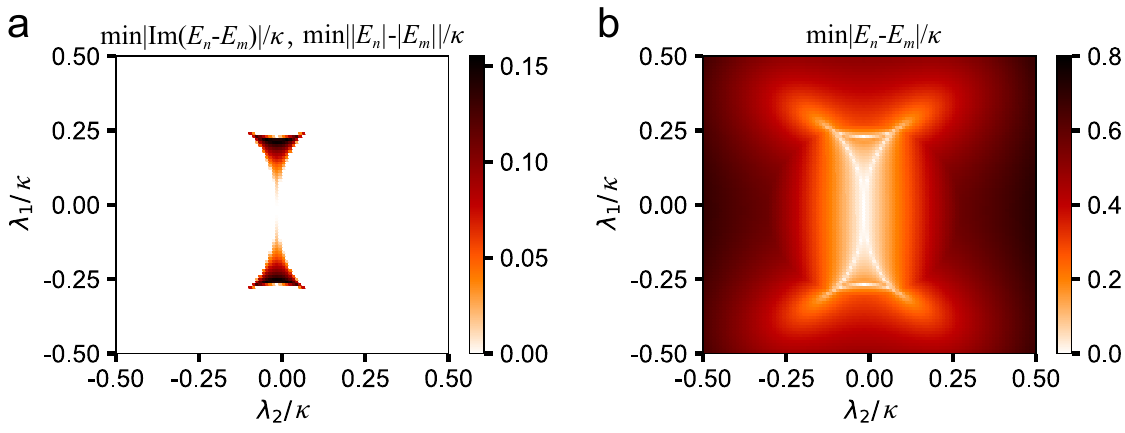}
	\caption{Characterization of eigenenergies. (a) The colored region is complementary to Fig.~1b in the main text. Outside this region, there is $\min|{\rm Im}(E_n-E_m)|=\min||E_n|-|E_m||=0$, indicating that at least two eigenenergies have equal imaginary parts and opposite real parts. (b) The lines of 2EPs are highlighted, which corresponds to $\min|E_n-E_m|=0$.}
	\label{min_dE}
\end{figure*}

In the single-excitation subspace \{$\left\vert 1_{1}0_{2}0_{3}\right\rangle
,\left\vert 0_{1}1_{2}0_{3}\right\rangle ,\left\vert
0_{1}0_{2}1_{3}\right\rangle $\}, $H$ has three eigenenergies, given by 
\begin{equation}
E_{1,2} = -\frac{i\kappa}{6} - \frac{1}{3}\left[ \left( -\frac{1}{2} \mp i\frac{\sqrt{3}}{2} \right) \frac{\xi}{\alpha} + \left(-\frac{1}{2} \pm i\frac{\sqrt{3}}{2}\right) \alpha \right],
\label{E1,2}
\end{equation}
and
\begin{equation}
E_3 = -\frac{i\kappa}{6} - \frac{1}{3}\left( \frac{\xi}{\alpha}+\alpha \right) , 
\label{E3}
\end{equation}
where 
\begin{equation}
\begin{aligned}
\xi &= 3\lambda^2_1 + 3\lambda^2_2 - \frac{\kappa^2}{4},\\
\alpha &= \sqrt[3]{\eta + \sqrt{\eta^2-\xi^3}},
\end{aligned}
\label{alpha}
\end{equation}
and
\begin{equation}
\eta = - \frac{i\kappa}{4}\left(18\lambda^2_2-9\lambda^2_1+\frac{\kappa^2}{2}\right).
\end{equation}
Figure \ref{min_dE}(a) illustrates either $\min|\mathrm{Im}(E_{n}-E_{m})|$ or $\min||E_n|-|E_m||$ versus $\lambda_1$ and $\lambda_2$ scaled in unit of $\kappa$. Outside the colored region, $\min|\mathrm{Im}(E_{n}-E_{m})|$ and $\min||E_n|-|E_m||$ are both $0$. Considering this in conjunction with Fig. 1b of the main text, it can be concluded that there are at least two eigenenergies with equal imaginary parts and opposite real parts. Therefore, outside this colored region, the eigenenergy can be expressed as
\begin{equation}
\begin{aligned}
E_1 &= iI_1, \\
E_2 &= R +iI , \\
E_3 &= -R +iI, \\
\end{aligned}
\end{equation}
where $R$, $I$, and $I_{1}$ are real parameters.
Figure \ref{min_dE}(b) shows the minimum of the scaled gaps $\min
\left\vert E_{n}-E_{m}\right\vert/\kappa $ ($n,m=1$ to $3$, $n\neq m$) among three
complex eigenenergies versus $\lambda _{1}$ and $\lambda _{2}$. The curves
with $\min \left\vert E_{n}-E_{m}\right\vert =0$ correspond to lines of
EP2s, where two of the three eigenenergies coalesce. The corresponding
eigenstates are
\begin{equation}
\left\vert \Phi _{j}\right\rangle =N_j\left\{\left[E_j\left(E_j+\frac{i\kappa}{2}\right)\right]|0_1 0_2 1_3\rangle + \lambda_2\left(E_j+\frac{i\kappa}{2}\right)|0_1 1_2 0_3\rangle + \lambda_1\lambda_2|1_1 0_2 0_3\rangle \right\}
\quad (j = 1,2,3), 
\end{equation}
where $N_j$ is the normalization factor.

When $\lambda _{1}\neq 0$ and $\lambda _{2}\neq 0$, each of these
eigenstates is a \textcolor{black}{tripartite} entangled state, for which each \textcolor{black}{mode} is entangled
with the other two \textcolor{black}{modes}. The system has four 3EPs at $\left\{ \pm \lambda
_{1}^{c},\pm \lambda _{2}^{c}\right\} $ with $\lambda _{1}^{c}=\sqrt{2}\kappa /3\sqrt{3}$ and $\lambda _{2}^{c}=\kappa /6\sqrt{3}$. For the EP3
in the first quadrant, the three-fold degenerate eigenenergy is
\begin{equation}
E_{{\rm EP3}} = -\frac{i\kappa}{6}
\end{equation}
with the corresponding eigenstate
\begin{equation}
|\Phi_{{\rm EP3}}\rangle = -\sqrt{\frac{1}{6}}|0_1 0_2 1_3\rangle + i\sqrt{\frac{1}{2}}|0_1 1_2 0_3\rangle + \sqrt{\frac{1}{3}} |1_1 0_2 0_3\rangle.
\end{equation}
This three-fold degenerate eigenstate is a genuine tripartite entangled state, manifested by the non-zero pairwise concurrences ${\cal C}_{1,2}=\sqrt{\frac{2}{3}}$, ${\cal C}%
_{2,3}=\sqrt{\frac{1}{3}}$, and ${\cal C}_{1,3}=\sqrt{\frac{2}{9}}$.

\section{Simulation of conditional dynamics in the isofrequency region}

\begin{figure*}[htbp]
	\centering
	\includegraphics[width=7in]{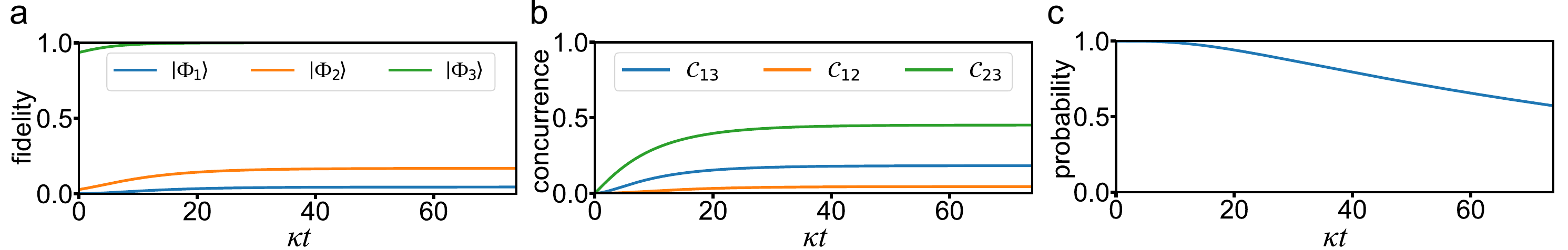}
	\caption{The fidelity (a), concurrence (b) and probability (c) evolution for the no-jump trajectory of the initial state $|0_10_21_3\rangle$ with $\lambda_1/\kappa=0.2$ and $\lambda_2/\kappa=0.02$. \textcolor{black}{$|\Phi_1\rangle $, $|\Phi_2\rangle $ and $|\Phi_3\rangle $ are the three eigenstates for which the coefficients of the components $|1_10_20_3\rangle $,  $|0_11_20_3\rangle $ and  $|0_10_21_3\rangle $ have a maximum modulus.}}
	\label{conc_fid_pro_FermiArc}
\end{figure*}

\color{black}
After some calculations, we find $\eta^2-\xi^3>0$ in the isofrequency region. Therefore, we can rewrite the parameter $\alpha$ in Eq.~\ref{alpha} as 
\begin{equation}
\alpha=\left(a+ib\right)^{1/3},
\end{equation}
where $a=\sqrt{\eta^2-\xi^3}$ and $b=-i\eta$ are real numbers. With this expression, it is easy to check
\begin{equation}
\begin{aligned}
|\alpha|^2 &=\left(a^2+b^2\right)^{1/3}\\
& = -\left(3\lambda_{1}^2+3\lambda_{2}^2-\frac{\kappa^2}{4}\right) \\
& = - \xi.
\end{aligned}
\end{equation}
Then, we obtain $\xi/|\alpha|^2=-1$. By using $1/\alpha=\alpha^{\ast}/|\alpha|^2$, we can further obtain that $\xi/\alpha=-\alpha^{\ast}$. Substituting the result into Eq.~\ref{E1,2} and Eq.~\ref{E3}, we have
\begin{equation}
E_{1,2} = - \frac{i\kappa}{6} - i \frac{1}{3} \left[ \pm\sqrt{3}\mathrm{Re}(\alpha)-\mathrm{Im}(\alpha) \right]
\end{equation}
and
\begin{equation}
E_3 = - \frac{i \kappa}{6} - i \frac{2}{3} \mathrm{Im}(\alpha).
\end{equation}
Obviously, the result shows that real parts of three eigenenergies vanish in the isofrequency region. Figure \ref{conc_fid_pro_FermiArc} shows the simulated evolutions of \textcolor{black}{corresponding eigenstates' fidelities, pairwise concurrences, and the probability for the no-jump trajectory of the  initial state $|0_10_21_3\rangle$ in the region} with $\lambda_1/\kappa=0.2$ and $\lambda_2/\kappa=0.02$ by using the NH Hamiltonian of Eq.~\ref{H_NH}. In such a region each of the pairwise concurrences tends to a fixed value for the no-jump case at the price of a progressively decreasing probability.

\color{black}

\section{Synthesis of the NH model}

\begin{figure*}[htbp]
	\centering
	\includegraphics[width=3in]{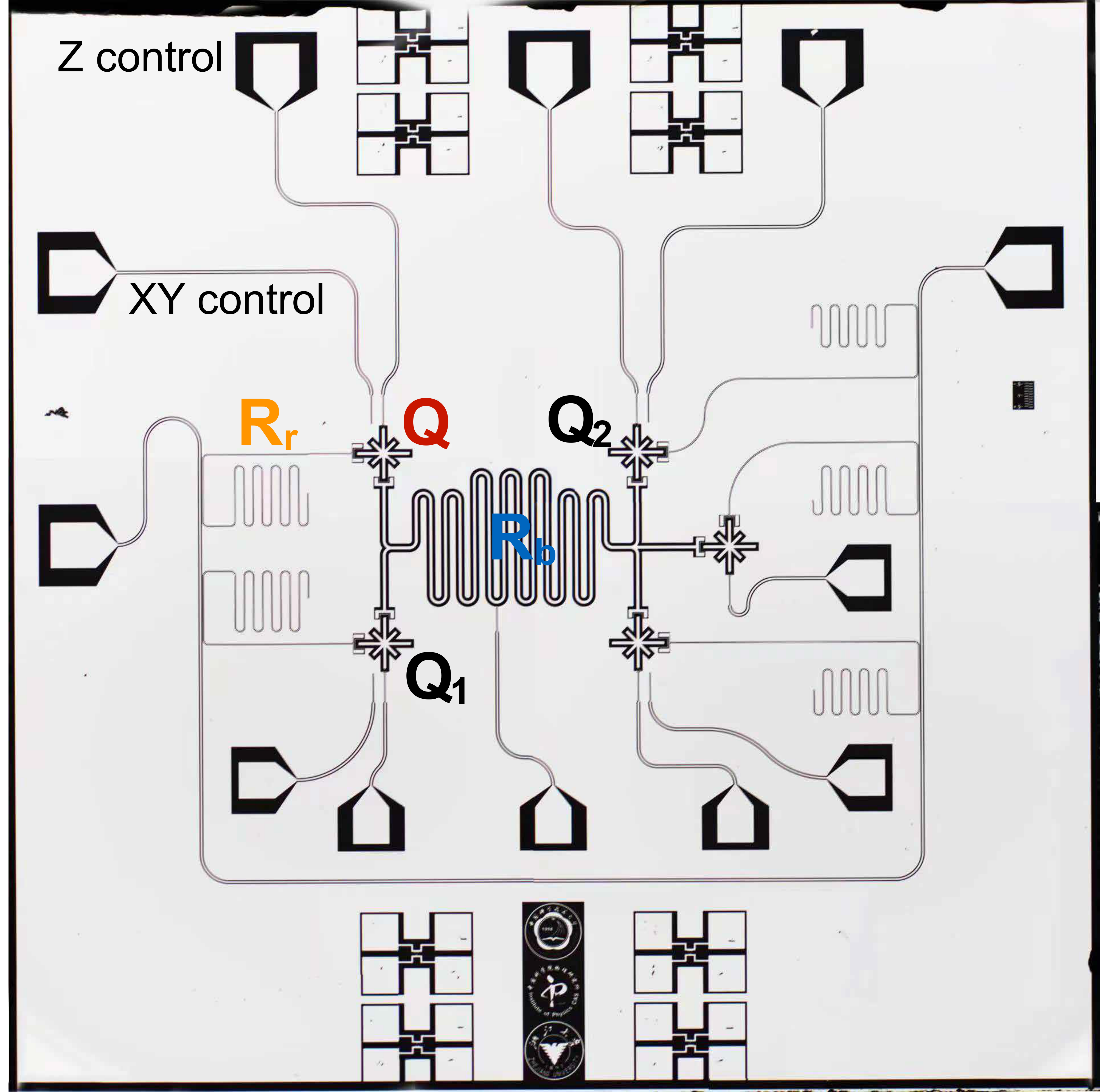}
	\caption{\textcolor{black}{Circuit micrography.}}
	\label{circuit}
\end{figure*}

The experiment is performed in a circuit quantum electrodynamics
architecture involving five frequency-tunable Xmon qubits, each individually
coupled to a readout resonator, and all connected to a bus resonator ($R_{b}$%
) with a fixed frequency $\omega _{b}/2\pi=5.58$ GHz, as sketched in Fig.~\ref{circuit}. 
\textcolor{black}{Every Xmon qubit used in our experiment has a microwave line (XY line) to drive its state transition and an individual flux line (Z line) to dynamically tune its frequency.}
The NH \textcolor{black}{three-mode system} is synthesized with the bus resonator, one of the
Xmon qubits ($Q$), and its readout resonator ($R_{r}$) with a fixed
frequency $\omega _{r}/2\pi=6.66$ GHz. 
The $R_{b}$-$Q$ and $Q$-$R_{r}$ swapping
interactions are realized by applying two parametric modulations to $Q$%
, making its frequency depend on time as%
\begin{equation}
\omega _{q}=\omega _{0}+\varepsilon _{1}\cos \left(\nu _{1}t\right)+\varepsilon
_{2}\cos \left(\nu _{2}t\right), 
\end{equation}%
where $\omega _{0}$ is the mean frequency, and $\varepsilon _{j}$ and $\nu
_{j}$ ($j=1,2$) are the corresponding modulation amplitude and angular
frequency of the $j$th modulation, respectively. 
\textcolor{black}{In the experiment, $\varepsilon _{j}$ and $\nu_{j}$ can be readily manipulated by a Z control line.}

$Q$ is capacitively coupled to $R_b$ and $R_r$ (see Ref. \cite{1} for details).
The coherent Hamiltonian of
the total system is given by%
\begin{equation}
H=H_{0}+H_{I}, 
\end{equation}%
where%
\begin{equation}
H_{0}=\omega _{b}a_{b}^{\dagger }a_{b}+\omega _{r}a_{r}^{\dagger
}a_{r}+\omega _{q}\left\vert e\right\rangle \left\langle e\right\vert , 
\end{equation}%
and%
\begin{equation}
H_{I}=\left\vert 0_{q}\right\rangle \left\langle 1_{q}\right\vert
\left(g_{b}a_{b}^{\dagger }+g_{r}a_{r}^{\dagger }\right)+\text{H.c.}, 
\end{equation}%
where $a_{r}^{\dagger }$ ($a_{b}^{\dagger }$) and $a_{r}$ ($a_{b}$) denote
the creation and annihilation operators for the photonic field stored in $%
R_{r}$ ($R_{b}$), $\left\vert 0_{q}\right\rangle $ and $\left\vert
1_{q}\right\rangle $ denote the ground and first excited states of $Q$, and $%
g_{b}$ ($g_{r}$) is the on-resonance $R_{b}$-$Q$ ($Q$-$R_{r}$) coupling
strength.

Performing the transformation $e^{i\int_{0}^{t}H_{0}dt}$, we obtain the
system Hamiltonian in the interaction picture,%
\begin{equation}
H_{I}^{\prime }=e^{-i\mu _{1}\sin \left(\nu _{1}t\right)}e^{-i\mu _{2}\sin \left(\nu
_{2}t\right)}\left\vert 0_{q}\right\rangle \left\langle 1_{q}\right\vert
\left(e^{i\delta _{b}t}g_{b}a_{b}^{\dagger }+e^{i\delta _{r}t}g_{r}a_{r}^{\dagger
}\right)+\text{H.c.}, 
\end{equation}%
where $\mu _{j}=\varepsilon _{j}/\nu _{j}$ ($j=1,2$), $\delta _{b}=\omega
_{b}-\omega _{0}$, and $\delta _{r}=\omega _{r}-\omega _{0}$. Using the
Jacobi-Anger expansion

\begin{equation}
e^{-i\mu _{j}\sin \left(\nu _{j}t\right)}=\stackrel{\infty }{%
\mathrel{\mathop{\sum }\limits_{n=-\infty }}%
}J_{n}(\mu _{j})e^{-in\nu _{j}t}, 
\end{equation}%
with $J_{n}(\mu _{j})$ being the $n$th Bessel function of the first kind, $%
H_{I}^{\prime }$ can be rewritten as%
\begin{equation}
H_{I}^{\prime }=\stackrel{\infty }{%
\mathrel{\mathop{\sum }\limits_{m,n=-\infty }}%
}J_{m}(\mu _{1})J_{n}(\mu _{2})e^{-im\nu _{1}t}e^{-in\nu _{2}t}\left\vert
0_{q}\right\rangle \left\langle 1_{q}\right\vert \left(e^{i\delta
_{b}t}g_{b}a_{b}^{\dagger }+e^{i\delta _{r}t}g_{r}a_{r}^{\dagger }\right)+\text{%
H.c.} 
\end{equation}%
Under the conditions $\nu _{1}=\delta _{r}$ and $\nu _{2}=-\delta _{b}$, $Q$
is resonantly coupled to $R_{r}$ ($R_{b}$) at the first upper (lower)
sideband with respect to the first (second) modulation. When $g_{b},g_{r}$ $%
\ll \nu _{1},\nu _{2}$, the fast-oscillating terms can be discarded, so that 
$H_{I}^{\prime }$ reduces to%
\begin{equation}
H_{I}^{\prime }=\left\vert 0_{q}\right\rangle \left\langle 1_{q}\right\vert
\left(\lambda _{1}a_{b}^{\dagger }+\lambda _{2}a_{r}^{\dagger }\right)+\text{H.c.}, 
\end{equation}%
where $\lambda _{1}=g_{r}J_{1}(\mu _{1})J_{0}(\mu _{2})$ and $\lambda
_{2}=g_{b}J_{0}(\mu _{1})J_{1}(\mu _{2})$, as
depicted in \textcolor{black}{Fig. 2a of the main text}. In the single-excitation
subspace, $a_{b}^{\dagger }$ and $a_{r}^{\dagger }$ can be replaced by $%
\left\vert 1_{b}\right\rangle \left\langle 0_{b}\right\vert $ and $%
\left\vert 1_{r}\right\rangle \left\langle 0_{r}\right\vert $, respectively.
Then $H_{I}^{\prime }$ is equivalent to the Hermitian part of the
Hamiltonian (1) of the main text with $N=3$. In our system, the dissipation
rates of $R_{b}$ and $Q$ are respectively $0.08$ MHz and $0.06$ MHz, which are negligible
compared with that of $R_{r}$. With the dissipation being included, the NH
Hamiltonian is given by Eq. \ref{H_NH}, with $R_{r}$, $Q$, and $R_{b}$
corresponding to the first, second, and last qubits, respectively.


\section{State readout}

The readout of the output $R_{b}$-$Q$-$R_{r}$ state is enabled with two
ancilla qubits, denoted as $Q_{1}$ and $Q_{2}$. After the NH Hamiltonian
dynamics, the state of $R_{b}$ is mapped to $Q_{1}$ through a swapping gate,
which is realized by tuning the transition frequency of $Q_{1}$ to $\omega
_{b}$ for a duration $t_{sw}=\pi /(2g_{1})\simeq 12.3$ $ns$, with $g_{1}=2\pi\times20.3$ MHz being
the $R_{b}$-$Q$ photonic swapping rate. Then, the state of $Q$ is transferred
to $Q_{2}$ by subsequently performing the $Q$-$R_{b}$ and $R_{b}$-$Q_{2}$
swapping gates. Finally, $R_{r}$'s state is transferred to $Q$. As the
maximum frequency of $Q$ ($2\pi\times6.01$ GHz) is smaller than $\omega _{r}$ by an amount much
lager than $g_{r}$, it is necessary to use the parametric modulation to
realize the $R_{r}$-$Q$ mapping. The corresponding gate duration is $150$ $ns$. With
a correction for the state distortion during the state mapping, the
resulting $Q_{1}$-$Q_{2}$-$Q$ output state corresponds to the $R_{b}$-$Q$-$%
R_{r}$ output state right before the state mapping.

\section{Extraction of eigenenergies}

\begin{figure*}[htbp]
	\centering
	\includegraphics[width=5in]{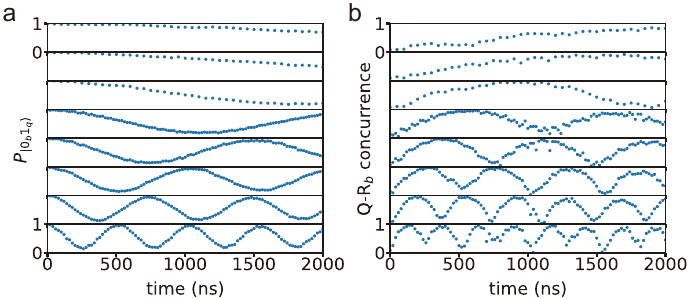}
	\caption{The population (a) and concurrence (b) evolutions for different values of $\lambda_2$ when $\lambda_1 = 0$, i.e., the edge from $(0,0)$ to $(0,\lambda_m)$. \textcolor{black}{From top to bottom, the values of $\lambda_2/2\pi$ are 0.04, 0.05, 0.14, 0.23, 0.33, 0.48, 0.68 and 0.97 MHz, respectively.}}
	\label{edge2_P_C}
\end{figure*}

In our experiment, we choose a square-shaped loop on the $\lambda _{1}$-$%
\lambda _{2}$ plane to extract the winding number. The four vertices of the
rectangle are ($0,0$), ($\lambda _{m},0$), ($0,\lambda _{m}$), and ($\lambda
_{m},\lambda _{m}$) with $\lambda _{m}\simeq2\pi\times1$ MHz. Along the edge with $\lambda
_{1}=0$, $R_{r}$ is decoupled from the $Q$-$R_{b}$ subsystem. In the
interaction picture, the $Q$-$R_{b}$ swapping coupling is described by the
Hamiltonian
\begin{equation}
{\cal H}=\lambda _{2}\left(a_{b}^{\dagger }\left\vert 0_{q}\right\rangle
\left\langle 1_{q}\right\vert +a_{b}\left\vert 1_{q}\right\rangle
\left\langle 0_{q}\right\vert \right), 
\end{equation}%
where $a_{b}^{\dagger }$ and $a_{b}$ denote the creation and annihilation
operators for the photonic mode stored in $R_{b}$, and $\left\vert
0_{q}\right\rangle $ and $\left\vert 1_{q}\right\rangle $ represent \textcolor{black}{the ground and excited states of $Q$}. In the single-excitation subspace, this
Hamiltonian has two eigenenergies $E_{\pm }=\pm \lambda _{2}$. The
corresponding eigenstates are 
\begin{equation}
\left\vert \Phi _{\pm }\right\rangle =\frac{1}{\sqrt{2}}\left( \left\vert
0_{b}1_{q}\right\rangle \pm \left\vert 1_{b}0_{q}\right\rangle \right) . 
\end{equation}%
The subsystem, starting from the initial state $\left\vert
0_{b}1_{q}\right\rangle $, evolves as%
\begin{equation}
\cos \left(\lambda _{2}t\right)\left\vert 0_{b}1_{q}\right\rangle -i\sin \left(\lambda
_{2}t\right)\left\vert 1_{b}0_{q}\right\rangle . 
\end{equation}%
The value of $\lambda _{2}$, which depends on the amplitude and frequency of
the parametric modulation used to mediate the sideband interaction, is
inferred from the observed Rabi oscillation. The population evolutions for
the state $\left\vert 0_{b}1_{q}\right\rangle $, observed for different
values of $\lambda _{2}$, are presented in Fig.~\ref{edge2_P_C}(a). The $Q$-$R_{b}$
concurrences associated with the two eigenstates $\left\vert \Phi _{\pm
}\right\rangle $, extracted at $\lambda _{2}=\lambda _{m}$, are 0.997 and 0.997,
respectively.

\begin{figure*}[htbp]
	\centering
	\includegraphics[width=7in]{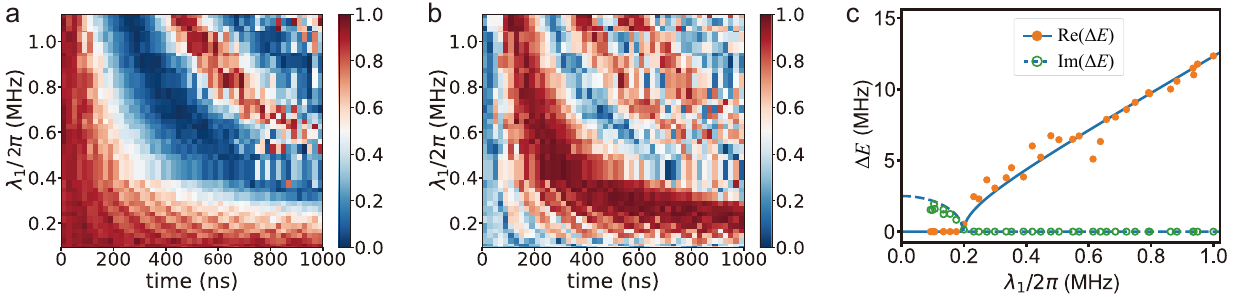}
	\caption{The population (a) and concurrence (b) evolutions for different values of $\lambda_1$ when $\lambda_2 = 0$, i.e., the edge from $(0,0)$ to $(\lambda_m,0)$. (c) Spectral gap $\Delta E$. The solid and dashed lines denote the real and imaginary parts, respectively.}
	\label{edge1_P_C_E}
\end{figure*}

For $\lambda _{2}=0$, $R_{b}$ is decoupled from the $Q$-$R_{r}$ subsystem.
In this case, the evolution of the $Q$-$R_{r}$ subsystem associated with the
no-jump trajectory is described by the NH Hamiltonian 
\begin{equation}
{\cal H}^{\prime }=\lambda _{1}\left(a_{r}^{\dagger }\left\vert
0_{q}\right\rangle \left\langle 1_{q}\right\vert +a_{r}\left\vert
1_{q}\right\rangle \left\langle 0_{q}\right\vert \right)-\frac{i}{2}\kappa
a_{r}^{\dagger }a_{r}, 
\end{equation}%
where $a_{r}^{\dagger }$ ($a_{r}$) is the photonic creation
(annihilation) operator for $R_{r}$. In the single-excitation subspace, the $%
{\cal H}^{\prime }$ has two eigenenergies 
\begin{equation}
E_{\pm }^{\prime }=-i\kappa /4\pm \sqrt{\lambda _{1}^{2}-\kappa ^{2}/16}. 
\end{equation}%
The corresponding eigenstates are 
\begin{equation}
\left\vert \Phi _{\pm }^{\prime }\right\rangle ={\cal N}_{\pm }\left(
\left\vert 1_{q}0_{r}\right\rangle +\frac{E_{\pm }^{\prime }}{\lambda _{1}}%
\left\vert 0_{q}1_{r}\right\rangle \right) , 
\label{eigenS}
\end{equation}%
where ${\cal N}_{\pm }=\left( 1+\left\vert E_{\pm }^{\prime }/\lambda
_{1}\right\vert ^{2}\right) ^{-1/2}$. Fig.~\ref{edge1_P_C_E}(a) shows the measured population
of the state $\left\vert 1_{q}0_{r}\right\rangle $ versus $\lambda _{1}$ and 
$t.$ This population is obtained by discarding the outcome $\left\vert
0_{q}0_{r}\right\rangle ${\sl , }and then renormalizing the probabilities
for the outcomes of $\left\vert 1_{q}0_{r}\right\rangle $ and $\left\vert
0_{q}1_{r}\right\rangle $. The gap of the two eigenenergies versus $\lambda _{1}$,
extracted from the population evolution, are shown in Fig.~\ref{edge1_P_C_E}(c).

When $\lambda _{1}\neq 0$ and $\lambda _{2}\neq 0$, the evolution of the $%
R_{b}$-$Q$-$R_{r}$ system associated with the no-jump trajectory is governed
by the Hamiltonian

\begin{equation}
{\cal H}={\cal H}^{H}+{\cal H}^{NH}, 
\end{equation}%
where 
\begin{equation}
\begin{aligned}
{\cal H}^{H} &=\left\vert 1_{q}\right\rangle \left\langle 0_{q}\right\vert
\left(\lambda _{1}a_{r}+\lambda _{2}a_{b}\right)+\text{H.c.}, \\
{\cal H}^{NH} &=-\frac{1}{2}i\kappa a_{r}^{\dagger }a_{r}.
\end{aligned} 
\end{equation}%
Suppose that the loop on the $\lambda _{1}-\lambda _{2}$ plane is a
rectangle with four vertices ($0,0$), ($\lambda _{m},0$), ($0,\lambda _{m}$%
), and ($\lambda _{m},\lambda _{m}$). On the edges from ($0,0$) to ($\lambda
_{m},0$) and to ($0,\lambda _{m}$), the system reduces to a \textcolor{black}{two-mode
system}, so that the eigenspectra can be extracted relatively easily. Without
dissipation, the three eigenstates of ${\cal H}^{H}$ on the other two edges
are given by

\begin{equation}
\begin{aligned}
\left\vert \Phi _{1}\right\rangle &=\sin \theta \left\vert
1_{b}0_{q}0_{r}\right\rangle -\cos \theta \left\vert
0_{b}0_{q}1_{r}\right\rangle , \\
\left\vert \Phi _{2}\right\rangle &=\frac{1}{\sqrt{2}}(\cos \theta
\left\vert 1_{b}0_{q}0_{r}\right\rangle +\sin \theta \left\vert
0_{b}0_{q}1_{r}\right\rangle +\left\vert 0_{b}1_{q}0_{r}\right\rangle ), \\
\left\vert \Phi _{3}\right\rangle &=\frac{1}{\sqrt{2}}(\cos \theta
\left\vert 1_{b}0_{q}0_{r}\right\rangle +\sin \theta \left\vert
0_{b}0_{q}1_{r}\right\rangle -\left\vert 0_{b}1_{q}0_{r}\right\rangle ),
\end{aligned} 
\end{equation}%
where $\tan \theta =\lambda_{1}/\lambda_{2}$. The corresponding
eigenenergies are $E_{1}^{H}=0$ and $E_{2}^{H}=-E_{3}^{H}=\lambda $, with $%
\lambda =\sqrt{\lambda _{1}^{2}+\lambda _{2}^{2}}$.

In the basis $\left\{ \left\vert \Phi _{j}\right\rangle \right\} $ ($j=1,2,3$), ${\cal H}^{NH}$ can be expressed as
\begin{equation}
{\cal H}^{NH}={\cal H}^{dg}+{\cal H}^{ndg}, 
\end{equation}%
where ${\cal H}^{dg}$ and ${\cal H}^{ndg}$ represent the diagonal and
off-diagonal parts respectively, given by%
\begin{equation}
{\cal H}^{dg}=\left( 
\begin{tabular}{lll}
$-i\frac{\kappa }{2}\cos ^{2}\theta $ & $0$ & $0$ \\ 
$0$ & $-i\frac{\kappa }{4}\sin ^{2}\theta $ & $0$ \\ 
$0$ & $0$ & $-i\frac{\kappa }{4}\sin ^{2}\theta $%
\end{tabular}%
\right) , 
\end{equation}%
and%
\begin{equation}
{\cal H}^{ndg}=\left( 
\begin{array}{ccc}
0 & i\frac{\kappa }{4\sqrt{2}}\sin \left(2\theta \right) & i\frac{\kappa }{4\sqrt{2}}%
\sin (2\theta ) \\ 
i\frac{\kappa }{4\sqrt{2}}\sin \left(2\theta \right) & 0 & i\frac{\kappa }{4}\sin
^{2}\theta \\ 
i\frac{\kappa }{4\sqrt{2}}\sin \left(2\theta \right) & i\frac{\kappa }{4}\sin ^{2}\theta
& 0%
\end{array}%
\right) . 
\end{equation}%
In the basis{\bf \ }$\left\{ \left\vert \Phi _{j}\right\rangle \right\} $,
we can rewrite the total Hamiltonian as%
\begin{equation}
{\cal H}={\cal H}^{0}+{\cal H}^{ndg}, 
\end{equation}%
where%
\begin{equation}
\begin{aligned}
{\cal H}^{0} &={\cal H}^{H}+{\cal H}^{dg} \\
&=\left( 
\begin{array}{ccc}
-i\frac{\kappa }{2}\cos ^{2}\theta & 0 & 0 \\ 
0 & \lambda -i\frac{\kappa }{4}\sin ^{2}\theta & 0 \\ 
0 & 0 & -\lambda -i\frac{\kappa }{4}\sin ^{2}\theta%
\end{array}%
\right) .
\end{aligned} 
\end{equation}

We note ${\cal H}^{ndg}$ can be treated as a perturbation, which is explained
as follow. On the edge with $\lambda _{1}=\lambda _{m}$, $\theta $ changes
from $\pi /2$ to $\pi /4$. For $\theta =\pi /2$, the non-zero off-diagonal
elements are ${\cal H}_{2,3}^{ndg}$ and $H_{3,2}^{ndg}$, which have a magnitude of $%
\left\vert {\cal H}_{2,3}^{ndg}\right\vert =\kappa /4$. The ratio of this
magnitude to the gap between the last two eigenvalues of ${\cal H}^{0}$ is $%
\left\vert {\cal H}_{2,3}^{ndg}\right\vert /\left\vert {\cal H}_{2,2}^{0}-%
{\cal H}_{3,3}^{0}\right\vert =\kappa /8\lambda\simeq 0.099$. When $%
\theta $ changes to $\pi /4$, $\left\vert {\cal H}_{2,3}^{ndg}\right\vert
/\left\vert {\cal H}_{2,2}^{0}-{\cal H}_{3,3}^{0}\right\vert $ monotonously
decreases to 0.070, while $\left\vert {\cal H}_{1,2}^{ndg}\right\vert
/\left\vert {\cal H}_{1,1}^{0}-{\cal H}_{2,2}^{0}\right\vert $ approximately
increases to $\kappa /(4\sqrt{2}\lambda )\simeq 0.099$. On the edge with $%
\lambda _{2}=\lambda_m$, $\theta $ changes from $0$ to $\pi /4$. For $%
\theta =0$, all the off-diagonal elements are 0. When $\theta $ increases to 
$\pi /4$, $\left\vert {\cal H}_{1,2}^{ndg}\right\vert /\left\vert {\cal H}%
_{1,1}^{0}-{\cal H}_{2,2}^{0}\right\vert $ and $\left\vert {\cal H}%
_{2,3}^{ndg}\right\vert /\left\vert {\cal H}_{2,2}^{0}-{\cal H}%
_{3,3}^{0}\right\vert $ approximately increase to $0.098$ and $0.099$, respectively.
These results imply that the magnitude of each off-diagonal element is much
smaller than the corresponding energy gap, which ensures the perturbation
condition. To the first order correction, the three eigenenergies correspond
to the diagonal elements of ${\cal H}^{0}$. This indicates that the real
part of the first eigenenergy $E_{1}$ is approximately zero, and the
other two eigenenergies $E_{2}$ and $E_{3}$ have the same
imaginary part but opposite real parts. Therefore, the three eigenenergies
can be approximately expressed as 
\begin{equation}
\begin{aligned}
E_{1} &\simeq -iI_{1}, \\
E_{2} &\simeq R-iI_{2}, \\
E_{3} &\simeq -R-iI_{2},
\end{aligned} 
\end{equation}
where $R$, $I_{1}$, and $I_{2}$ are real parameters, with $I_{1}=\frac{%
\kappa }{2}\cos ^{2}\theta $ and $I_{2}=\frac{\kappa }{4}\sin ^{2}\theta $.
Consequently, the eigenenergies are determined by these three parameters $%
\{R $, $I_{1}$, $I_{2}\}$, which can be extracted through observation of the
population evolutions.

\begin{figure*}[htbp]
	\centering
	\includegraphics[width=7in]{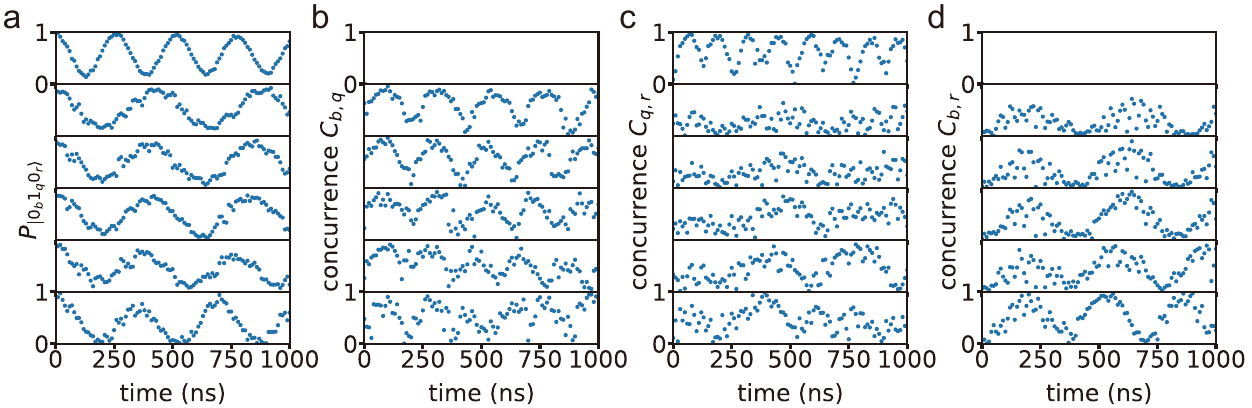}
	\caption{The population (a) and concurrence (b),(c),(d) evolutions for different values of $\lambda_1$ when $\lambda_2 = \lambda_m$, i.e., the edge from $(0,\lambda_m)$ to $(\lambda_m,\lambda_m)$. \textcolor{black}{From top to bottom, the values of $\lambda_1/2\pi$ are 0, 0.49, 0.58, 0.77, 0.83 and 1.01 MHz, respectively.}}
	\label{edge3_P_C}
\end{figure*}

\begin{figure*}[htbp]
	\centering
	\includegraphics[width=5in]{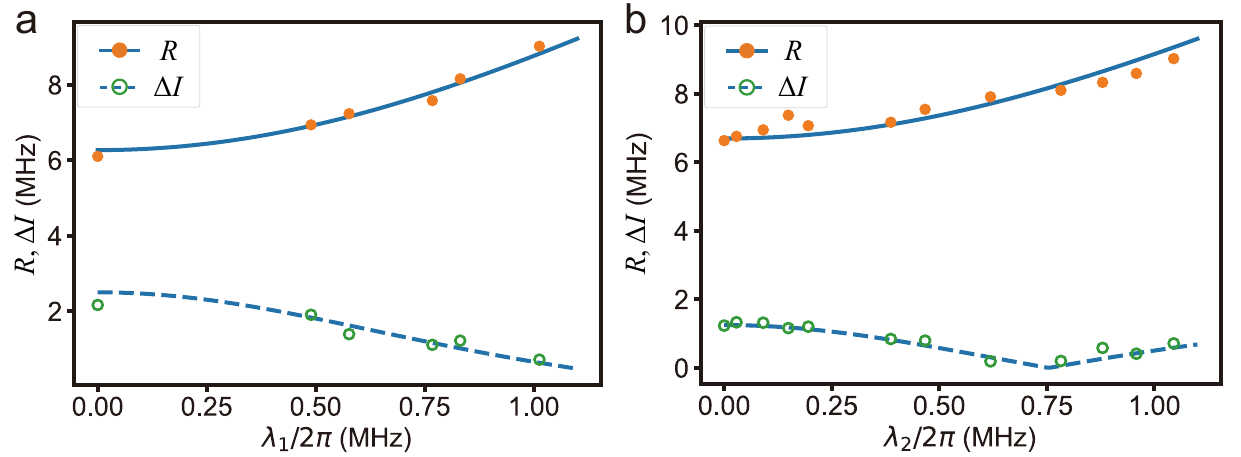}
	\caption{The extracted eigenenergies in terms of $R$ and $\Delta I$ versus (a) $\lambda _{1}$ and (b) $\lambda_2$. }
	\label{eigenE_R_I}
\end{figure*}

Figure~\ref{edge3_P_C}(a) displays the measured population of the state $\left\vert
0_{b}1_{q}0_{r}\right\rangle $ versus $\lambda _{1}$ and $t$ for the edge
with $\lambda _{2}=\lambda _{m}$. This population is obtained by discarding
the outcome $\left\vert 0_{b}0_{q}0_{r}\right\rangle $, and then
renormalizing the probabilities of the three single-excitation outcomes. The eigenenergies in terms of $R$ and $\Delta I$ ($=|I_1-I_2|$) versus $\lambda _{1}$, extracted from this population
evolution, are displayed in Fig.~\ref{eigenE_R_I}(a). Figure~\ref{edge4_P_C}(a) shows
the measured $\left\vert 0_{b}1_{q}0_{r}\right\rangle $-state population
versus $\lambda _{2}$ and $t$ for the edge with $\lambda _{1}=\lambda _{m}$.
The extracted eigenenergies in terms of $R$ and $\Delta I$ versus $\lambda _{2}$ are displayed in
Fig.~\ref{eigenE_R_I}(b).

\section{Characterization of the nonclassicality}

\begin{figure*}[htbp]
	\centering
	\includegraphics[width=7in]{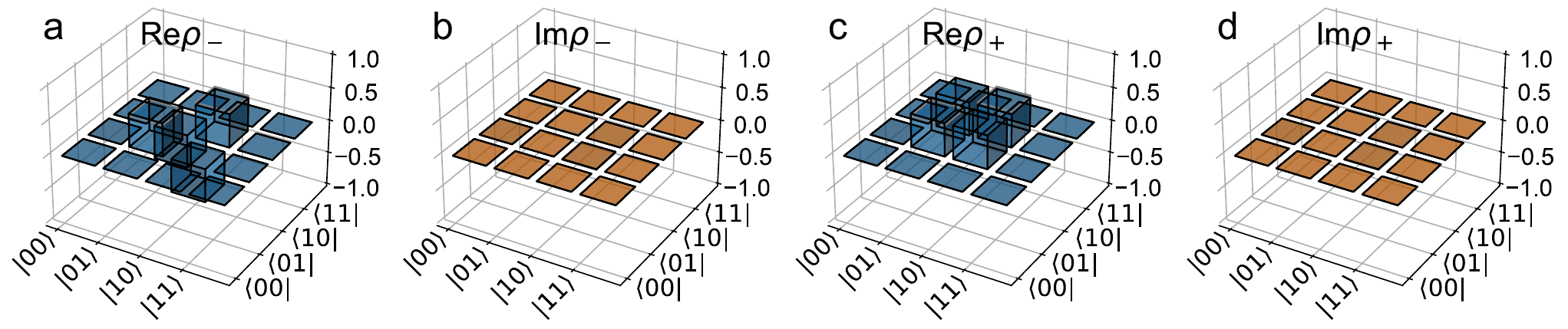}
	\caption{The density matrices $\rho_{\pm}$ for the two eigenstates $\left\vert \Phi _{\pm }\right\rangle$. (a) The real parts of $\rho_-$. (b) The imaginary parts of $\rho_-$. (c) The real parts of $\rho_+$. (d) The imaginary parts of $\rho_+$. The two numbers in each ket denote the excitation numbers of the qubit and the bus resonator, respectively. The black frames denote the matrix elements of the ideal eigenstates.} 
	\label{eigenstate_QRb}
\end{figure*}

\begin{figure*}[htbp]
	\centering
	\includegraphics[width=7in]{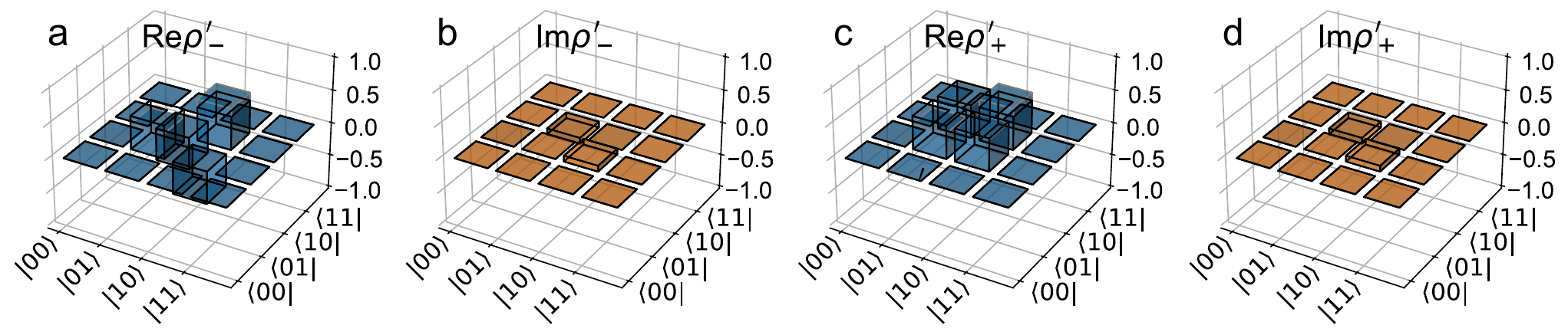}
	\caption{The density matrices $\rho_{\pm}^{\prime}$ for the two eigenstates $\left\vert \Phi _{\pm }^{\prime}\right\rangle$. (a) The real parts of $\rho_-^{\prime}$. (b) The imaginary parts of $\rho_-^{\prime}$. (c) The real parts of $\rho_+^{\prime}$. (d) The imaginary parts of $\rho_+^{\prime}$. The two numbers in each ket denote the excitation numbers of the qubit and the readout resonator, respectively. The black frames denote the matrix elements of the ideal eigenstates.}
	\label{eigenstate_QRr}
\end{figure*}

Along the edge with $\lambda _{1}=0$, $R_{r}$ remains in
the ground state $\left\vert 0_{r}\right\rangle $. The measured $Q$-$R_{b}$ concurrence, versus $\lambda _{2}$ and $t$, is displayed in Fig.~\ref{edge2_P_C}(b). The density matrices for the two eigenstates $\left\vert \Phi _{\pm }\right\rangle $, extracted from the data
measured at $\lambda _{2}=\lambda _{m}$, are presented in Fig.~\ref{eigenstate_QRb}. The concurrences corresponding to these two eigenstates are 0.997 and 0.997,
respectively. Fig.~\ref{edge1_P_C_E}(b) shows the $Q$-$R_{r}$ concurrence versus $\lambda _{1}$ and $t$, measured for the edge $\lambda _{2}=0$. The measured density matrices, associated with the two eigenstates $
\left\vert \Phi _{\pm }^{\prime}\right\rangle $ for $\lambda
_{1}=\lambda _{m}$, are presented in Fig.~\ref{eigenstate_QRr}, with the concurrences 0.971
and 0.971.

\begin{figure*}[htbp]
	\centering
	\includegraphics[width=7in]{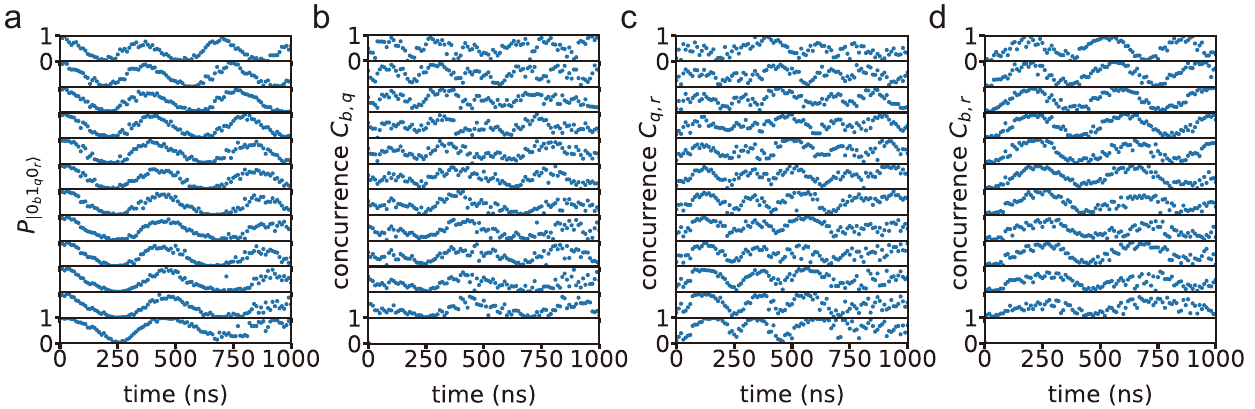}
	\caption{The population (a) and concurrence (b),(c),(d) evolutions for different values of $\lambda_2$ when $\lambda_1 = \lambda_m$, i.e., the edge from $(\lambda_m,0)$ to $(\lambda_m,\lambda_m)$. \textcolor{black}{From top to bottom, the values of $\lambda_2/2\pi$ are 1.04, 0.96, 0.88, 0.78, 0.62, 0.47, 0.39, 0.20, 0.15, 0.09, 0.03 and 0 MHz, respectively.}}
	\label{edge4_P_C}
\end{figure*}

Figures \ref{edge3_P_C}(b), (c), and (d) present the measured three pairwise concurrences $C_{b,q}$, $C_{q,r}$, and $C_{b,r}$ versus $\lambda _{1}$ and $t$ for the edge with $\lambda _{2}=\lambda
_{m}$, where the subscript ``$b$'', ``$q$'', and ``$r$'' denote the bus resonator, Xmon qubit, and readout resonator, respectively.
Figure~\ref{edge4_P_C}(b), (c), and (d) showcase the three pairwise concurrences versus $\lambda_{2}$ and $t$ for the edge with $\lambda_{1}=\lambda_{m}$. These results show that the tripartite system evolves from the initial product state $\left\vert 0_{b}1_{q}0_{r}\right\rangle $ to a tripartite entangled state under the NH Hamiltonian when $\lambda_{1}\neq 0$ and $\lambda _{2}\neq 0$. For example, the three concurrences, measured at the point with $\lambda _{1}=\lambda_{2}=\lambda _{m}$ for the time $600$ $ns$, are $C_{b,q}=0.50$, $C_{q,r}=0.64$, and $C_{b,r}=0.74$, respectively. These results imply that the
corresponding eigenstates are highly-nonclassical states, featuring tripartite quantum entanglement.

\section{Derivation of the resultant vector}
The resultant is a basic concept in algebra. It can be used to determine whether two polynomials have common roots, defined as
\begin{equation}
R_{P_1,P_2} \equiv {\rm{det}} S_{P_1,P_2},
\end{equation}
where $P_1$, $P_2$ are two polynomials and $S_{P_1,P_2}$ is their Sylvester matrix. Suppose
\begin{equation}
\begin{aligned}
P_1 &= a_0 x^n + a_1 x^{n-1} + ... +a_n,\\
P_2 &= b_0 x^m + b_1 x^{m-1} + ... +b_m, 
\end{aligned}
\end{equation}
the corresponding resultant is a determinant of order $m + n$,
\begin{equation}
R_{P_1,P_2} = 
\begin{vmatrix}
a_0 & a_1 & a_2 & ... & ... & a_n & 0 & ... & 0 \\
0 & a_0 & a_1 & ... & ... & a_{n-1} & a_n & ... & 0\\
\vdots & \vdots & \vdots & \vdots & \vdots & \vdots & \vdots & \vdots & \vdots \\
0 & 0& ... & 0 & a_0 & ... & ... & ... & a_n \\
b_0 & b_1 & b_2 & ... & ... & ... & b_m & ... &0 \\
0 & b_0 & b_1  & ... & ... & ... & b_{m-1} & b_m & ... \\
\vdots & \vdots & \vdots & \vdots & \vdots & \vdots & \vdots & \vdots & \vdots \\
0 & ... & 0 & b_0 & b_1 & ... & ... & ... & b_m

\end{vmatrix}.
\end{equation}
If $R_{P_1,P_2} = 0$, the polynomials $P_1$ and $P_2$ have common roots.

For a three-dimensional system governed by the Hamiltonian $H$, the characteristic polynomial $P$ is given by
\begin{equation}
P = -(E-E_1)(E-E_2)(E-E_3),
\end{equation}
where $E_1$, $E_2$ and $E_3$ are the three eigenvalues of $H$. The first- and second-order derivatives of $P(E)$ are
\begin{equation}
P^{\prime}=-[(E-E_1)(E-E_2)+(E-E_2)(E-E_3)+(E-E_3)(E-E_1)],
\end{equation}
and
\begin{equation}
P^{\prime\prime}=-2[(E-E_1)+(E-E_2)+(E-E_3)].
\end{equation}
The two components of the resultant vector are given by
\begin{equation}
\mathcal{R}_1=-R_{P,P^{\prime}}=(E_1-E_2)^2(E_1-E_3)^2(E_2-E_3)^2
\end{equation}
and
\begin{equation}
\mathcal{R}_2=iR_{P,P^{\prime\prime}}=8(E_1+E_3-2E_2)(E_1+E_2-2E_3)(E_2+E_3-2E_1).
\end{equation}
Therefore, we can calculate ${\cal R}_{1}$ and ${\cal R}_{2}$ at each point ($\lambda _{1},\lambda _{2}$%
) of the parameter space with the measured eigenenergies. The results are presented in Fig. 3 of the main text.

\section{Extraction of the winding number}

\begin{figure*}[htbp]
	\centering
	\includegraphics[width=6in]{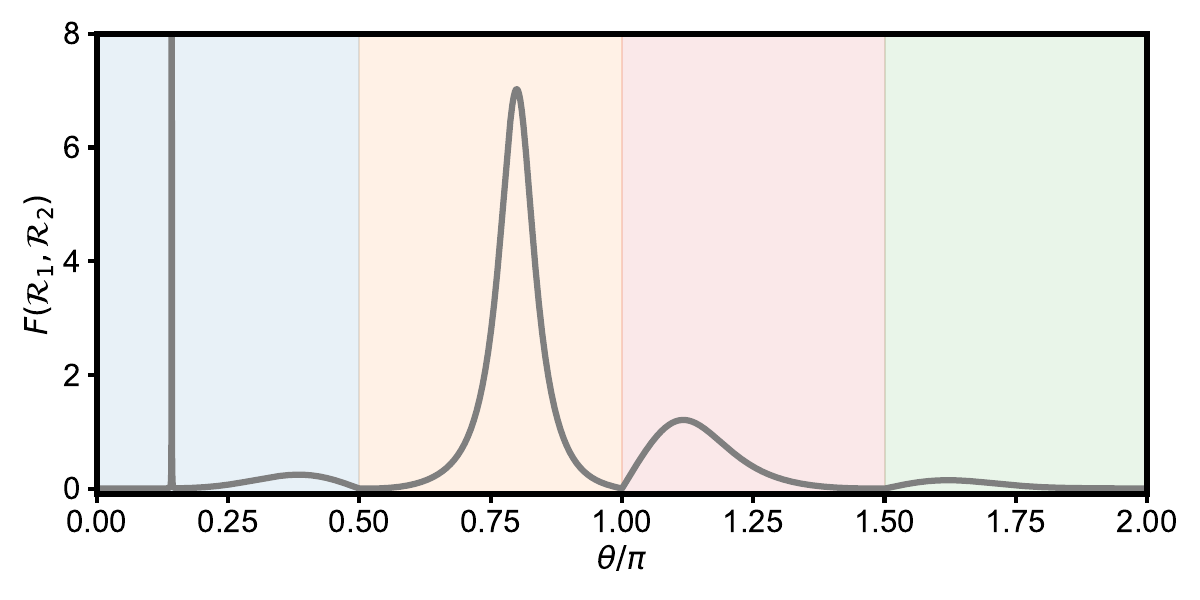}
	\caption{$F({\cal R}_{1},{\cal R}_{2})$ along the four edges. Areas with different background colors correspond to the four edges.}
	\label{dW}
\end{figure*}

The winding number, associated with each EP3, is calculated along a loop enclosing the EP3,
\begin{equation}
{\cal W}=\frac{1}{2\pi}\mathop{\displaystyle\sum}\limits_{j=1,2}\displaystyle\oint\limits_{C_{\lambda }}F({\cal R}_{1},{\cal R}_{2})d\lambda _{j}, 
\end{equation}
where the integrand is given by%
\begin{equation}
F({\cal R}_{1},{\cal R}_{2})=\frac{1}{\left\Vert {\cal R}\right\Vert ^{2}}%
\left( {\cal R}_{1}\frac{\partial {\cal R}_{2}}{\partial \lambda _{j}}-{\cal %
R}_{2}\frac{\partial {\cal R}_{1}}{\partial \lambda _{j}}\right) . 
\end{equation}%
The square-shaped loop chosen in our experiment encloses the EP3 in the
first quadrant. Only one control
parameter changes along each edge of the loop. Thus the integral can be rewritten as%
\begin{equation}
\begin{aligned}
{\cal W} &=&\frac{1}{2\pi}\int_{0}^{\lambda _{m}}F({\cal R}_{1},{\cal R}_{2})d\lambda
_{1}|_{\lambda _{2}=0} \\
&&+\frac{1}{2\pi}\int_{0}^{\lambda _{m}}F({\cal R}_{1},{\cal R}_{2})d\lambda
_{2}|_{\lambda _{1}=\lambda _{m}} \\
&&+\frac{1}{2\pi}\int_{\lambda _{m}}^{0}F({\cal R}_{1},{\cal R}_{2})d\lambda
_{1}|_{\lambda _{2}=\lambda _{m}} \\
&&+\frac{1}{2\pi}\int_{\lambda _{m}}^{0}F({\cal R}_{1},{\cal R}_{2})d\lambda
_{2}|_{\lambda _{1}=0},
\end{aligned}
\end{equation}
where $F({\cal R}_{1},{\cal R}_{2})$ along the four edges are
displayed in Fig.~\ref{dW}.

For simplicity, the square-shaped trajectory can be represented by the parametric equation
\begin{equation}
\begin{aligned}
\lambda_1&= \frac{1}{2}\left(1-\cos\theta|\cos\theta|+\sin\theta|\sin\theta|\right),\\
\lambda_2&=\frac{1}{2}\left(1-\cos\theta|\cos\theta|-\sin\theta|\sin\theta|\right),\\
\end{aligned}
\end{equation}
where $\theta$ ranges from $0$ to $2\pi$. In this case, the winding number in terms of $\theta$ is given by
\begin{equation}
{\cal W} = \frac{1}{2\pi}\int_{0}^{2\pi}
\frac{1}{\left\Vert {\cal R}\right\Vert ^{2}}
\left( {\cal R}_{1}\frac{\partial {\cal R}_{2}}{\partial \theta}-{\cal R}_{2}\frac{\partial {\cal R}_{1}}{\partial \theta}\right) d\theta.
\end{equation}

\section{Winding numbers for DPs and EP2s}

To confirm that the winding number, defined by Eq.3 of the main text, is indeed uniquely associated with the topology for the three-order EPs, we here calculate this quantity for the DPs (diabolical points) in both the two-dimensional (2D) and 3D Hermitian system, as well as for the EP2s in a 2D NH system. 

We first consider the 2D Hermitian system with the Hamiltonian $H = \omega_x \sigma_x + \omega_y \sigma_y$, where $\omega_x$ ($\omega_y$) is a real number.
This model admits two real eigenenergies, and one could study the DP at $\omega_x = \omega_y =0$. The characteristic polynomial of $H$ is
\begin{equation}
P(E) = E^2-(\omega_x^2+\omega_y^2),
\end{equation}
and its first-order derivative is
\begin{equation}
P^{\prime}(E) = 2E.
\end{equation}
The second-order derivative of $P(E)$ is a constant, $P^{\prime\prime}(E)=2$.
In order to construct the Sylvester matrix of $P(E)$ and $P^{\prime\prime}(E)$, we can take $P^{\prime\prime}(E)$ as a polynomial of degree $1$ with a zero coefficient, 
\begin{equation}
P^{\prime\prime}(E)=2 + 0E.
\end{equation}
In this way, we have
\begin{equation}
S_{P,P^{\prime}} = \left( 
\begin{array}{cccc}
1 & 0 & -(\omega_x^2+\omega_y^2) \\ 
2 & 0 & 0 \\
0 & 2 & 0
\end{array}
\right)
\text{ and }
S_{P,P^{\prime\prime}} = \left( 
\begin{array}{cccc}
1 & 0 & -(\omega_x^2+\omega_y^2) \\ 
0 & 2 & 0 \\
0 & 0 & 2
\end{array}
\right).
\end{equation}
The corresponding resultant $\mathcal{R}_1=R_{P,P^{\prime}}\equiv\det S_{P,P^{\prime}} =1-4(\omega_x^2+\omega_y^2)$ is a real-valued function, while $\mathcal{R}_2=R_{P,P^{\prime\prime}}\equiv\det S_{P,P^{\prime\prime}} =4$ is a constant. 
Therefore, along a closed loop around DP in a 2D space spanned by $\omega_x$ and $\omega_y$, the trajectory of resultant vector $\mathcal{R}$ in the $\mathcal{R}_1-\mathcal{R}_2$ plane is a line segment instead of a closed curve, 
which means the winding number is always $0$. We further note that the spectrum of any 2D Hermitian system has a similar structure, so that the corresponding winding number is zero. 

We now turn to the 3D Hermitian system, whose dynamics is described by the three-mode Hamiltonian of Eq.~\ref{H_NH} with $\kappa=0$. In the single-excitation subspace, the Hamiltonian takes the form
\begin{equation}
H = \left( 
\begin{array}{cccc}
0 & \lambda_2 & 0\\ 
\lambda_2 & 0 & \lambda_1 \\
0 & \lambda_1 & 0
\end{array}
\right),
\label{H3x3}
\end{equation}
with a DP at $\lambda_1=\lambda_2=0$.
The characteristic polynomial reads
\begin{eqnarray}
P(E)=-E^3+(\lambda_1^2+\lambda_2^2)E
\end{eqnarray}
and its derivatives are
\begin{eqnarray}
P^{\prime}(E)=-3E^2+(\lambda_1^2+\lambda_2^2),
\end{eqnarray}
\begin{eqnarray}
P^{\prime\prime}(E)=-6E.
\end{eqnarray}
We then have
\begin{equation}
S_{P,P^{\prime}} = \left( 
\begin{array}{ccccc}
-1 & 0 & \lambda_1^2+\lambda_2^2 & 0 & 0 \\ 
0 & -1 & 0 & \lambda_1^2+\lambda_2^2 & 0 \\
-3 & 0 & \lambda_1^2+\lambda_2^2 & 0 & 0 \\
0 & -3 & 0 & \lambda_1^2+\lambda_2^2 & 0 \\
0 & 0 & -3 & 0 & \lambda_1^2+\lambda_2^2
\end{array}
\right)
\end{equation}
and
\begin{equation}
S_{P,P^{\prime\prime}} = \left( 
\begin{array}{ccccc}
-1 & 0 & \lambda_1^2+\lambda_2^2 & 0 \\ 
-6 & 0 & 0 & 0 \\
0 & -6 & 0 & 0 \\
0 & 0 & -6 & 0 
\end{array}
\right).
\end{equation}
The corresponding resultants are
\begin{equation}
\begin{aligned}
\mathcal{R}_1&=R_{P,P^{\prime}}\equiv\det S_{P,P^{\prime}}\\
&= 4\lambda_2^6 + 12\lambda_2^4\lambda_1^2+12\lambda_2^2\lambda_1^4 + 4\lambda_1^6
\end{aligned}
\end{equation}
and
\begin{equation}
\begin{aligned}
\mathcal{R}_2&=R_{P,P^{\prime\prime}}\equiv\det S_{P,P^{\prime\prime}}\\
&= 0.
\end{aligned}
\end{equation}
Similar to \textit{Example 1}, $\mathcal{R}_1$ is a real-valued function, while $\mathcal{R}_2$ is a constant. Therefore, we reach the same conclusion: the winding number defined by Eq.3 of the main text for the third-order DP is $0$.

Finally, we consider the 2D dissipative system, with the NH Hamiltonian
\begin{equation}
H =\left( 
\begin{array}{cccc}
0 & J_x-iJ_y \\ 
J_x+iJ_y & -i\gamma/2 
\end{array}
\right),
\end{equation}
where $\gamma$ is the dissipation rate and $|J|=\sqrt{J_x^2+J_y^2}$ is the coupling strength. This Hamiltonian has two EP2s at $J=\pm\gamma/4$. The characteristic polynomial is
\begin{equation}
P(E)=E^2+i\gamma/2E-|J|^2,
\end{equation} 
and its derivatives read
\begin{equation}
P^{\prime}(E)=2E+i\gamma/2,
\end{equation} 
\begin{equation}
P^{\prime\prime}(E)=2.
\end{equation} 
From this, we can obtain the Sylvester matrices
\begin{equation}
S_{P,P^{\prime}} = \left( 
\begin{array}{cccc}
1 & i\frac{\gamma}{2} & -|J|^2 \\ 
2 & i\frac{\gamma}{2} & 0 \\
0 & 2 & i\frac{\gamma}{2} 
\end{array}
\right)
\text{ and }
S_{P,P^{\prime\prime}} = \left( 
\begin{array}{cccc}
1 & i\frac{\gamma}{2} & -|J|^2 \\ 
0 & 2 & 0 \\
0 & 0 & 2
\end{array}
\right).
\end{equation}
Subsequently, two components of the resultant vector $\mathbf{\mathcal{R}}$ are given by
\begin{equation}
\mathcal{R}_1=\gamma^2/4-4|J|^2,
\end{equation}
and
\begin{equation}
\mathcal{R}_2=4.
\end{equation}
Consistent with \textit{Example 1} and \textit{2}, $\mathcal{R}_1$ is a real-valued function, while $\mathcal{R}_2$ is a constant.
As a result, along a closed loop around each EP2 in the $J_x-J_y$ parameter space, the corresponding winding number of ${\bf \mathcal{R}}$ is $0$. 

These results confirm the claim of Ref. 12 of the main text that thus-defined winding number serves as a homotopy invariant, which uniquely characterizes the topology of EP3s.